%% file: HIG-12-005_temp.tex
\pdfoutput=1

\documentclass[11pt,twoside,a4paper,cmspaper,final,collab]{cms-tdr}
\def\svnVersion{177388}\def\cmsCernNo{2012-169}\def\cmsCernDate{\today}\def\cmsMessage{Submitted to the European Physical Journal C}

\begin{document}\cmsNoteHeader{HIG-12-005}

\hyphenation{had-ron-i-za-tion}
\hyphenation{cal-or-i-me-ter}
\hyphenation{de-vices}

\RCS$Revision: 149707 $
\RCS$HeadURL: svn+ssh://svn.cern.ch/reps/tdr2/papers/HIG-12-005/trunk/HIG-12-005.tex $
\RCS$Id: HIG-12-005.tex 149707 2012-09-28 13:34:11Z liis $
\newlength\cmsFigWidth
\ifthenelse{\boolean{cms@external}}{\setlength\cmsFigWidth{0.85\columnwidth}}{\setlength\cmsFigWidth{0.4\textwidth}}
\ifthenelse{\boolean{cms@external}}{\providecommand{\cmsLeft}{top}}{\providecommand{\cmsLeft}{left}}
\ifthenelse{\boolean{cms@external}}{\providecommand{\cmsRight}{bottom}}{\providecommand{\cmsRight}{right}}

\newcommand{\dblh}{\Phi^{++}}
\newcommand{\dbl}{\Phi}
\newcommand{\trip}{\dblh\Phi^{-}}
\newcommand{\tauhad}{\tau_\mathrm{h}}
\newcommand{\deltaphi}{\Delta\varphi}
\newcommand{\lumitot}{$4.93\pm0.11$ fb$^{-1}$}
\newcommand{\eeres}{444}
\newcommand{\emres}{453}
\newcommand{\mmres}{459}
\newcommand{\etres}{373}
\newcommand{\mtres}{375}
\newcommand{\ttres}{204}
\newcommand{\bponeres}{383}
\newcommand{\bptwores}{408}
\newcommand{\bpthreeres}{403}
\newcommand{\bpfourres}{400}
\newcommand{\eerespp}{382}
\newcommand{\emrespp}{391}
\newcommand{\mmrespp}{395}
\newcommand{\etrespp}{293}
\newcommand{\mtrespp}{300}
\newcommand{\ttrespp}{169}
\newcommand{\bponerespp}{333}
\newcommand{\bptworespp}{359}
\newcommand{\bpthreerespp}{355}
\newcommand{\bpfourrespp}{353}
\newcommand{\RelIso}{\ensuremath{\sum \text{RelIso}}\xspace}

\cmsNoteHeader{HIG-12-005} 
\title{A search for a doubly-charged Higgs boson in pp collisions at $\sqrt{s} = 7$\TeV}

\date{\today}

\abstract{
A search for a doubly-charged Higgs boson in pp collisions at $\sqrt{s} = 7$\TeV is presented.  The data correspond
to an integrated luminosity of 4.9\fbinv, collected by the CMS experiment at the LHC. The search is performed using events with three or more isolated charged leptons of any flavor, giving sensitivity to the decays of pair-produced triplet components $\Phi^{++}\Phi^{--}$, and $\Phi^{++}\Phi^{-}$ from associated production. No excess is observed compared to the background prediction, and
upper limits at the 95\% confidence level are set on the $\Phi^{++}$ production cross section, under specific assumptions on its branching fractions. Lower bounds on the $\Phi^{++}$ mass are reported, providing significantly more stringent constraints than previously published limits.}

\hypersetup{%
pdfauthor={CMS Collaboration},%
pdftitle={A search for a doubly-charged Higgs boson in pp collisions at sqrt(s) = 7 TeV},%
pdfsubject={CMS},%
pdfkeywords={CMS, physics, higgs, see-saw}}

\maketitle 
\ifthenelse{\boolean{cms@external}}{\sloppy}{}
\section{Introduction}

The existence of non-zero neutrino masses may represent a signal of physics beyond the standard model (SM)~\cite{PDG}.
The observation of a doubly-charged scalar particle would establish the type II seesaw mechanism as the most promising framework for generating neutrino masses~\cite{Raidal:2008jk}. The minimal type II seesaw model~\cite{Magg:1980ut, Schechter:1980gr,Lazarides:1980nt,Mohapatra:1980yp} is realized with an additional scalar field that is a triplet under $SU(2)_L$ and carries $U(1)_Y$ hypercharge $Y=2$. The triplet contains a doubly-charged component $\Phi^{++}$, a singly-charged component $\Phi^{+}$ and a neutral component $\Phi^{0}$. In this paper, the symbols $\dblh$ and $\Phi^+$ are used to refer also to the charge conjugate states $\Phi^{--}$ and $\Phi^-$. In the literature $\Delta$ and $H$ have also been used. Our choice of the symbol $\Phi$ for the triplet components avoids possible confusion with the minimal supersymmetric model (MSSM) $H^{+}$ boson.

The $\dblh$ particle carries double electric charge, and decays to same-sign lepton pairs $\ell_\alpha^+\ell_\beta^+$ with flavor indices $\alpha, \beta$, where $\alpha$ can be equal to or different from $\beta$. The $\Phi^{++}$ Yukawa coupling matrix $Y_\Phi$ is proportional to the light neutrino
mass matrix. The measurement of the $\Phi^{++}\to \ell_\alpha^+\ell_\beta^+$ branching fractions would therefore allow the neutrino mass generation mechanism to be tested~\cite{Hektor:2007uu}. In this scenario, measurements at the Large Hadron Collider (LHC) could shed light~\cite{Garayoa:2007fw,Kadastik:2007yd,Akeroyd:2007zv,Chun:2003ej} on the
absolute neutrino mass scale, the mass hierarchy, and the Majorana CP-violating phases. The latter are not measurable in current neutrino-oscillation experiments.

\begin{figure*}[hbtp]
  \begin{center}
    \includegraphics[width=0.45\textwidth]{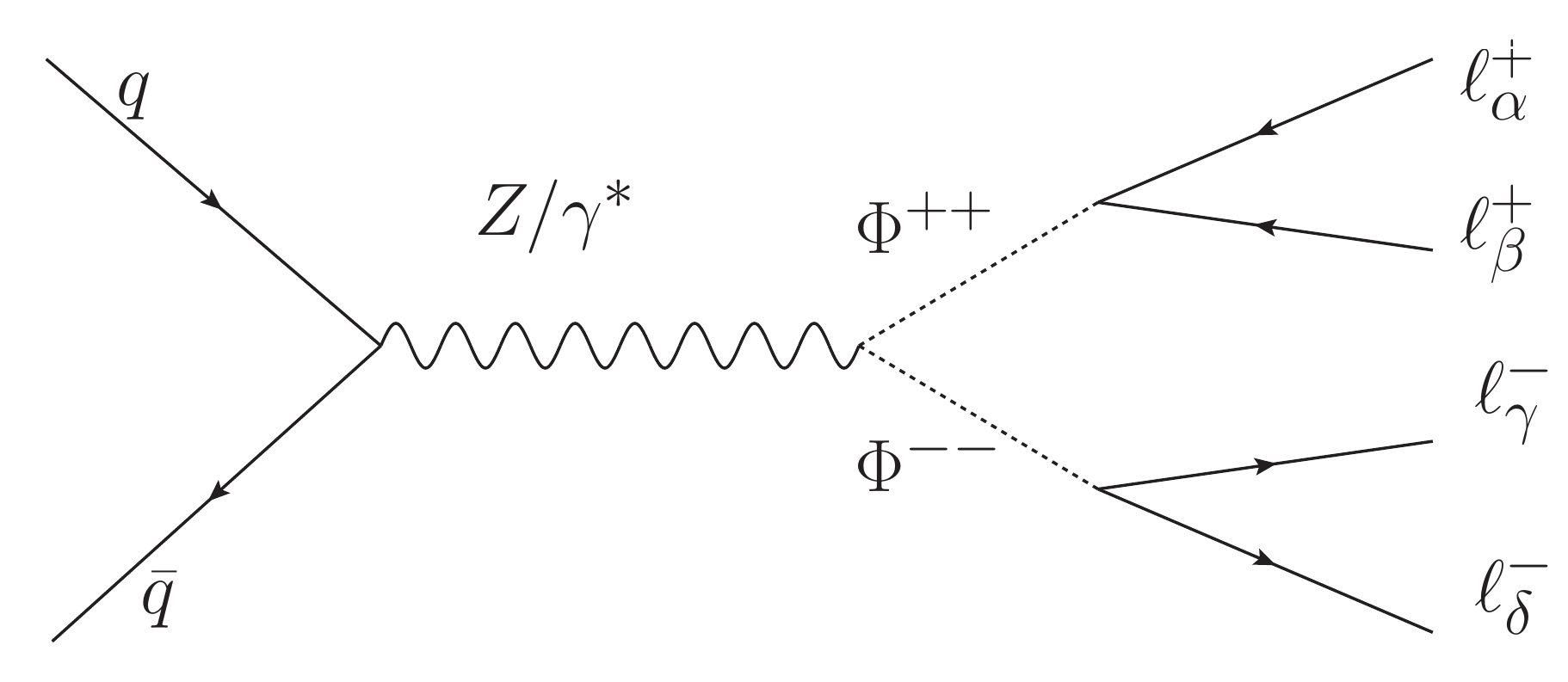}\includegraphics[width=0.45\textwidth]{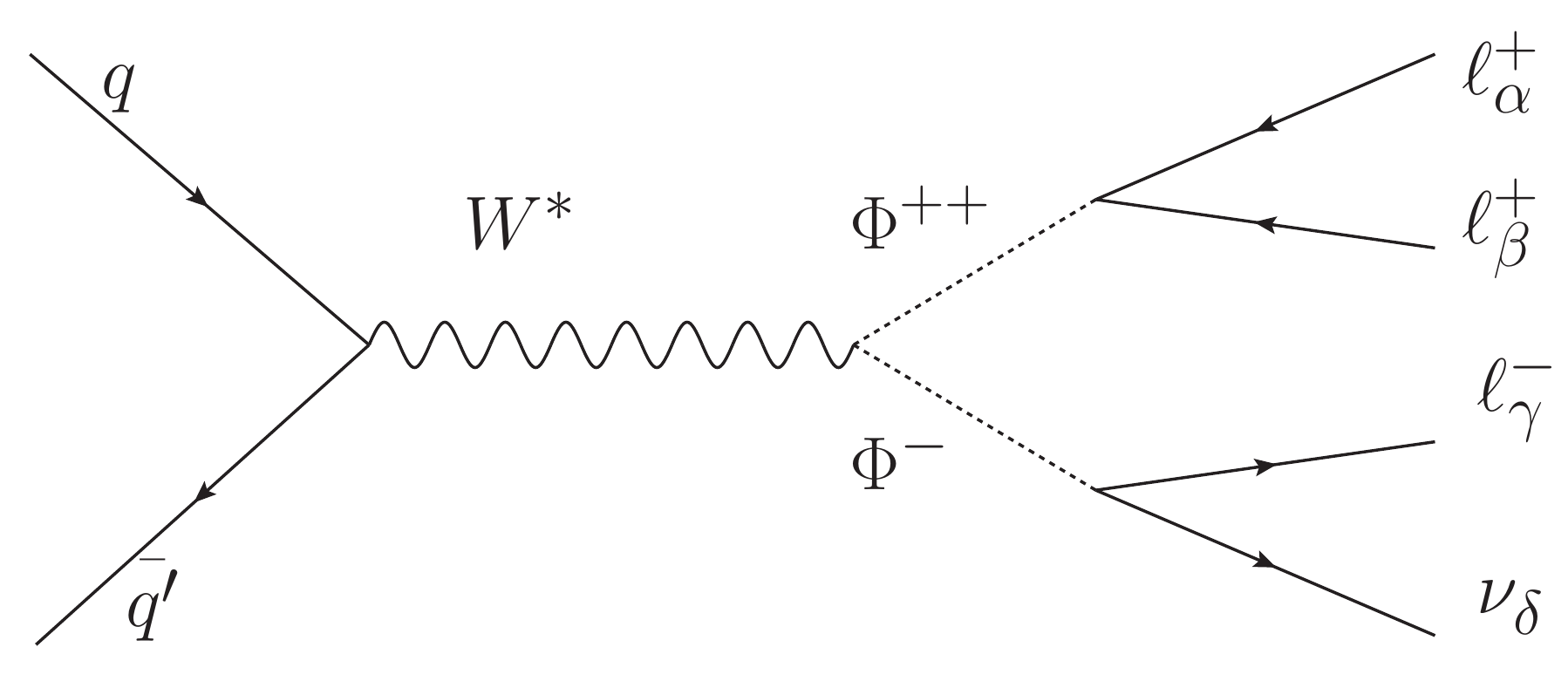}
        \caption{Feynman diagrams for pair and associated production of $\dblh$.}
    \label{fig:feynman}
  \end{center}
\end{figure*}

\begin{figure*}[hbtp]
  \begin{center}
    \includegraphics[width=\textwidth]{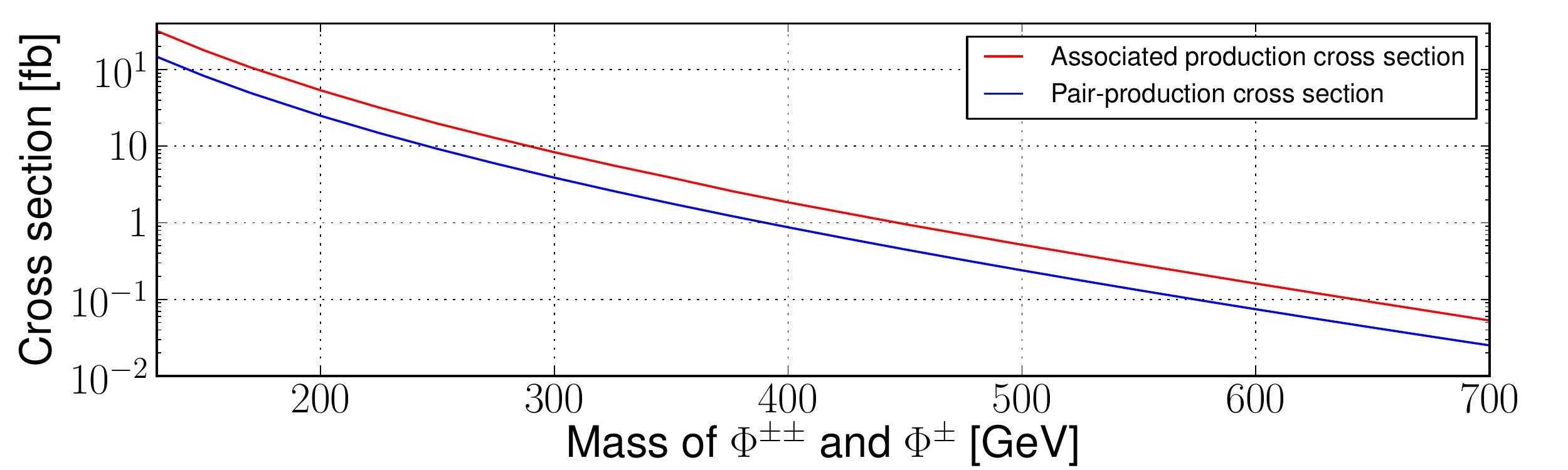}
        \caption{Production cross sections for pair and associated production processes at $\sqrt{s} = 7$\TeV.}
    \label{fig:xsec}
  \end{center}
\end{figure*}

In this article the results of an inclusive search for a doubly-charged Higgs boson at the Compact Muon Solenoid (CMS) experiment
are presented, based on a dataset corresponding to an integrated luminosity of \lumitot. The dataset was collected in pp collisions at $\sqrt{s} = 7$\TeV during the 2011 LHC running period. Both the pair-production process $\Pp\Pp \to \Phi^{++}\Phi^{--}\to \ell_\alpha^+\ell_\beta^+\ell_\gamma^-\ell_\delta^-$ \cite{Huitu:1996su,Muhlleitner:2003me} and the associated production process $\Pp\Pp\to \Phi^{++}\Phi^{-}\to  \ell_\alpha^+\ell_\beta^+\ell_\gamma^-\nu_\delta^{ }$ \cite{Akeroyd:2005gt,Akeroyd:2010ip} are studied. It is assumed that the $\dblh$ and $\Phi^{+}$ are degenerate in mass. However, as the singly-charged component is not fully reconstructed, this requirement impacts only the cross section, as long as the mass splitting is such that cascade decays (e.g. $\Phi^{++}\rightarrow\Phi^+\PW^{+*}\rightarrow\Phi^0 \PW^{+*}\PW^{+*}$) are disfavored~\cite{Melfo:2011nx}. The relevant Feynman diagrams and production cross sections, calculated following~\cite{Muhlleitner:2003me}, are presented in Figures~\ref{fig:feynman} and \ref{fig:xsec}. The $\Phi^{++}\to \PW^{+}\PW^{+}$ decays are assumed to be suppressed. In the framework of type II seesaw model~\cite{Magg:1980ut, Schechter:1980gr,Lazarides:1980nt,Mohapatra:1980yp}, where the triplet is used to explain neutrino masses, this is a natural assumption:
the decay width to the $\PW^{+}\PW^{+}$ channel is proportional to the vacuum expectation value of
the triplet ($v_\Phi$) and, as the neutrino masses are determined from the product of the Yukawa couplings and $v_\Phi$,
then large enough $v_\Phi$ values would require unnaturally small Yukawa couplings.

The search strategy is to look for an excess of events in one or more flavor combinations of same-sign lepton pairs
coming from the decays $\Phi^{++}\to \ell_\alpha^+\ell_\beta^+$. Final states containing three or four charged leptons
are considered.

In addition to a model-independent search in each final state, where the $\Phi^{++}$ is assumed to decay in 100\% of the cases
in turn in each of the possible lepton combinations ($\Pe\Pe, \mu \mu, \tau\tau, \Pe\mu, \Pe\tau, \mu\tau$), the type II seesaw model is tested, following~\cite{Kadastik:2007yd}, at four benchmark
points (BP), that probe different neutrino mass matrix structures.
BP1 and BP2 describe a neutrino sector with a massless neutrino, assuming normal and inverted mass hierarchies, respectively. BP3 represents a degenerate neutrino mass spectrum with the mass taken as 0.2\unit{eV}. The fourth
benchmark point BP4 represents the case in which the $\Phi^{++}$ has an equal branching fraction to each lepton generation. This corresponds to the following values of the Majorana phases: $\alpha_1=0$, $\alpha_2=1.7$. BP4 is the  only case in which $\alpha_2$ is non-vanishing. For all benchmark points,
vanishing CP phases and an exact tri-bimaximal neutrino mixing matrix are assumed, fixing the values of the
mixing angles at $\theta_{12}=\sin^{-1}(1/\sqrt{3})$, $\theta_{23}=\pi/4$, and $\theta_{13}=0$. The four benchmark points, along with the model-independent search,
encompass the majority of the parameter space of possible $\dblh$ leptonic decays. The values of the neutrino parameters
at the benchmark points are compatible with currently measured values within uncertainties. The recent measurement of a
non-zero $\theta_{13}$ angle~\cite{An:2012eh,Ahn:2012nd} is the only exception, and influences the branching fractions at the benchmark points by a maximum of a few percent~\cite{Kadastik:2007yd}. The branching fractions at the benchmark points are summarized in Table~\ref{tab:branchings}.

\begin{table*}[!htbp]
\begin{center}
\topcaption{Branching fractions of $\dblh$ at the four benchmark points. }
\label{tab:branchings}
\begin{tabular}{|c|c|c|c|c|c|c|}
\hline
Benchmark point & ee & e$\mu$ & e$\tau$ & $\mu\mu$ & $\mu\tau$ & $\tau\tau$ \\
\hline
BP1 & 0 & 0.01 & 0.01 & 0.30 & 0.38 & 0.30 \\
BP2 & 1/2 & 0 & 0 & 1/8 & 1/4 & 1/8 \\
BP3 & 1/3 & 0 & 0 & 1/3 & 0 & 1/3 \\
BP4 & 1/6 & 1/6 & 1/6 & 1/6 & 1/6 & 1/6 \\
\hline
\end{tabular}
\end{center}
\end{table*}

The first limits on the $\dblh$ mass were derived based on the measurements done at PEP and PETRA
experiments~\cite{Swartz:1989qz,Althoff:1983ig,Berger:1984kw,Derrick:1985xj,Derrick:1986de,Fernandez:1986dt}.
Next, the $\dblh$ was  searched for at the MARK II detector at SLAC~\cite{Swartz:1990ki}, the H1 detector at HERA \cite{Aktas:2006nu} and the LEP
experiments~\cite{Achard:2003mv,Abdallah:2002qj,Abbiendi:2003pr,Abbiendi:2001cr}. The latest results are from the Tevatron
and ATLAS~\cite{Aaltonen:2011rta,Aad:2012cg,Abazov:2011xx} experiments, which set lower limits on the $\dblh$ mass between 112 and 355\GeV,
depending on assumptions regarding $\dblh$ branching fractions. In all previous searches, only the pair-production mechanism, and only a small fraction of the possible final state combinations, were
considered. The addition of associated production and all possible final states significantly improves the sensitivity and reach of this analysis.

\section{The CMS detector}
The central feature of the CMS apparatus is a superconducting solenoid of 6\unit{m} internal diameter with a 3.8\unit{T} field. Within the field volume are a silicon pixel and strip tracker, a crystal electromagnetic calorimeter (ECAL) and a brass/scintillator hadron calorimeter. Muons are measured in gas-ionization detectors embedded in the steel return yoke. Extensive forward calorimetry complements the coverage provided by the barrel and endcap detectors.

CMS uses a right-handed coordinate system, with the origin at the nominal interaction point, the $x$ axis pointing to the center of the LHC ring, the $y$ axis pointing up (perpendicular to the LHC ring), and the $z$ axis along the counterclockwise-beam direction. The polar angle, $\theta$, is measured from the positive $z$-axis and the azimuthal angle, $\phi$, is measured in the $x$-$y$ plane.

The inner tracker measures charged particles within the pseudorapidity range $|\eta| < 2.5$, where $\eta=-\ln[\tan(\theta/2)]$. It consists of 1440 silicon pixel and 15\,148 silicon strip detector modules, and is located in the superconducting solenoid. It provides an impact parameter resolution of ${\sim}15\mum$ and a transverse momentum ($\pt$) resolution of about 1.5\% for 100\GeV particles.
The electromagnetic calorimeter consists of 75\,848 lead tungstate crystals which provide coverage in pseudorapidity $\vert \eta \vert< 1.479 $ in the barrel region and $1.479 <\vert \eta \vert < 3.0$ in two endcap regions (EE). A preshower detector consisting of two planes of silicon sensors interleaved with a total of three radiation lengths of lead is located in front of the EE.
The muons are measured in the pseudorapidity range $|\eta|< 2.4$, with detection planes made using three technologies: drift tubes, cathode strip chambers, and resistive plate chambers. Matching the muons to the tracks measured in the silicon tracker results in a transverse momentum resolution between 1 and 5\%, for $\pt$ values up to 1\TeV.
The detector is highly hermetic, ensuring accurate measurement
of the global energy balance in the plane transverse to the beam directions.

The first level of the CMS trigger system, composed of custom hardware processors, uses information from the calorimeters and muon detectors to select, in less than 1\mus, the most interesting events. The High Level Trigger processor farm further decreases the event rate from around 100\unit{kHz} to around 300\unit{Hz}, before data storage.
A detailed description of the CMS detector may be found in Reference~\cite{CMSdetector}.
\section{Experimental signatures}
The most important experimental signature of the $\dblh$ is the presence of two like-charge leptons in the final state,
with a resonant structure in their invariant mass spectrum. In this final state the background from SM processes is expected to be very small.
For the four-lepton final state from $\Phi^{++}\Phi^{--}$ pair production, both Higgs bosons may be reconstructed,
giving two like-charge pairs of leptons with similar invariant mass.

Like-charge backgrounds arise from various SM processes, including di-boson events containing two to four leptons in the final state. The Z+jets and $\ttbar$+jets, with leptonic W decays, contribute to the non-resonant background through jet misidentification as leptons, or via genuine leptons within jets. The W+jets and QCD multijet events are examples of large cross section processes which potentially contribute to the SM background. However, the requirement of multiple isolated leptons with high transverse momentum almost entirely removes the contribution from these processes.

\section{Monte Carlo simulations}\label{sec:MC}
The multi-purpose Monte Carlo (MC) event generator \PYTHIA 6.4.24~\cite{Sjostrand:2006za} is used for the simulation of signal and background processes, either to generate a given hard interaction at leading order (LO), or for the simulation of showering and hadronization in cases where the hard processes are generated at next-to-leading order (NLO) outside \PYTHIA, as in the case of top quark related backgrounds. The \TAUOLA~\cite{Jadach:1993hs} program is interfaced with \PYTHIA to simulate $\tau$ decay and polarization. Signal samples in the associated production mode are generated by using \textsc{CALCHEP} 2.5.2~\cite{Pukhov:2004ca}, as \PYTHIA only contains the doubly-charged particle. The diboson and Drell--Yan events are generated using \MADGRAPH 5.1.1.0~\cite{Alwall:2007st} and \TAUOLA. Samples of $\ttbar$+jets and single-top production are generated by using \POWHEG~\cite{Alioli:2009je,Nason:2004rx,Frixione:2007vw} and \PYTHIA.

The signal processes were simulated at 16 mass points: 130, 150, 170, 200, 225, 250, 275, 300, 325, 350, 375, 400, 450, 500, 600 and 700\GeV.
\section{Event selection}
\subsection{Trigger}
Collision events are selected through the use of double-lepton (ee, e$\mu$, $\mu\mu$) triggers.
In the case of the ee and e$\mu$ triggers, a minimum \pt of 17 and 8\GeV is required of the two leptons respectively.
In the case of the $\mu\mu$ trigger, the muon \pt thresholds changed during the data-taking period because of
the increasing instantaneous luminosity. A 7\GeV \pt threshold was applied to each muon during the initial
data-taking period (the first few hundred \pbinv). The thresholds were later raised to 13 and 8\GeV
for the two muons, and then to 17 and 8\GeV. The trigger efficiency is in excess of 99.5\% for the events
passing the selection defined below.

\subsection{Lepton identification}

The electron identification uses a cut-based approach in order to reject jets misidentified as electrons, or electrons originating from photon conversions.  Electron candidates are separated into categories according to the
amount of emitted bremsstrahlung energy; the latter depends on the  magnetic field intensity and the large and varying amount of material in front of the electromagnetic calorimeter. A bremsstrahlung recovery procedure creates superclusters (i.e. groups of clusters), which collect the energy released both by the electron and the emitted photons. Transverse energy ($\ET$) dependent and $\eta$-dependent selections are applied~\cite{EGM-10-004}.

Selection criteria for electrons include: geometrical matching between the position of the energy deposition in the ECAL
and the direction of the corresponding electron track; requirements on shower shape; the impact parameter of the electron track; isolation of the electron; and further selection criteria to reject photon conversions. To reduce contamination in the signal region, electrons must pass a triple charge determination procedure based on two different track curvature fitting algorithms and on the angle between the supercluster and the pixel hits. In addition, electrons are required to have $\pt>15$\GeV and $|\eta|<2.5$.

Muon candidates are reconstructed using two algorithms. The first matches tracks in the silicon detector to segments in the muon chambers, whereas the second performs a combined fit using hits in both the silicon tracker and the muon systems~\cite{MUO-10-002}. All muon candidates are required to be successfully reconstructed by both algorithms, and to have $\pt > 5\GeV$ and $|\eta|< 2.4$.

Isolation of the final state leptons plays a key role in suppressing backgrounds from $\ttbar$ and Z+jets.
A relative isolation variable (RelIso) is used, defined as the sum of the \pt of the tracks in the tracker and the energy
from the calorimeters in an isolation cone of size 0.3 around the lepton, excluding the contribution of the lepton candidate itself,
divided by the lepton \pt. A typical LHC bunch-crossing at high instantaneous luminosity results in overlapping proton-proton
collisions (`pileup'). The isolation variable is corrected for energy deposition within the isolation cone by pile-up events,
by means of the {\sc FastJet} energy-density algorithm~\cite{Cacciari:2007fd,Cacciari:2008gn}. A description of the performance
of the isolation algorithm in collision data can be found in~\cite{MUO-10-002,EGM-10-004}.

In order to reconstruct hadronic $\tau$ candidates ($\tauhad$), the `hadron plus strips' (HPS) algorithm~\cite{TAU-11-001} is used, which is based on particle flow (PF)~\cite{PFT-10-002} objects. One of the main tasks in reconstructing hadronically-decaying $\tau$ is determining the number of $\pi^0$ mesons produced in the decay. The HPS method combines PF electromagnetic objects
into `strips' at constant $\eta$ to take into account the broadening of calorimeter deposits due to conversions of $\pi^0$ decay photons.
The neutral objects are then combined with charged hadrons to reconstruct the $\tauhad$ decay.

The $\tauhad$ candidates are required to have $\pt> 15\GeV$ and $|\eta|<2.1$. Additional criteria
are applied to discriminate against e and $\mu$, since these particles could be misidentified as one-prong $\tauhad$. The $\tauhad$ candidates in the region $1.460 < |\eta| < 1.558$ are vetoed, owing to the reduced ability to discriminate between electrons and hadrons in the barrel-to-endcap transition region.

In the following, the term lepton is used to indicate both light leptons (e, $\mu$) and the $\tau$-lepton
before decay ($\tau$). It is not possible to distinguish between leptonic $\tau$ decay products and prompt
light leptons. Therefore, in scenarios that include a $\tau$ the light lepton contribution is assumed to be
a mixture of prompt and non-prompt particles and selection criteria are tuned accordingly. Beyond that
there is no attempt to distinguish the origin of the light leptons. As a result, a final state
$\Pep \Pep \tauhad^-$ could arise from $\dblh\Phi^{-}\rightarrow \Pep \Pep \tau^- \Pgngm\rightarrow \Pep \Pep\tauhad^- \Pagngt \Pgngm$ as well as from $\dblh\Phi^{-}\rightarrow \Pep \tau^+ \tau^- \Pgngm\rightarrow \Pep\Pep\nu_\tau \Pagne \tauhad^- \Pagngt \Pgngm$.
In both scenarios we look for a resonance in the $\Pep\Pep$ invariant mass, which is narrow in the case of direct signal decay to light-leptons and wide in the case of the presence of a $\tau$ in the intermediate state. Because of the reconstruction efficiency we treat the $\mathcal{ B}(\dblh\to\tau^+\tau^+)=100\%$ assumption separately and optimize the selection criteria accordingly.
However a given event may be assigned to more than one signal type if it matches the corresponding final state
(the above mentioned example event would contribute to all scenarios where e$\tau$, $\tau\tau$ branching fractions are non-zero assuming the event passes the respective selection criteria).

\subsection{Pre-selection requirements and signal selection optimization method}
\label{sec:preselection}
In order to select events from well-measured collisions, a primary vertex pre-selection is applied, requiring the number of
degrees of freedom for the vertex fit to be greater than 4, and the distance of the vertex from the center of the CMS
detector to be less than 24 cm along the beam line, and less than 2 cm in the transverse plane. In case of multiple primary vertex candidates,
the one with the highest value of the scalar sum of the total transverse momentum of the associated tracks is selected~\cite{TRK-10-005}.

Data and simulated events are preselected by requiring at least two final-state light leptons, with $\pt > 20\GeV$ and $\pt > 10\GeV$ respectively. If pairs of light leptons with invariant mass less than 12\GeV are reconstructed, neither of the particles is considered in the subsequent steps of the analysis. This requirement rejects low-mass resonances and light leptons from B meson decays. In order to reduce the background contribution from QCD multijet production and misidentified leptons, the two least well-isolated light leptons are required to have summed relative isolation ($\RelIso$) less than 0.35. In case of the $\mathcal{B}$$(\dblh\rightarrow\tau^+\tau^+)=100\%$ assumption, the requirement is tightened to less than 0.25.

In addition, the significance of the impact parameter, $\mathrm{SIP}_\ell = \rho_\mathrm{PV}/\Delta\rho_\mathrm{PV}$,
is required to be less than four for the reconstructed light leptons except for the $\mathcal{B}(\dblh\rightarrow\tau^+\tau^+)=100\%$ assumption; here $\rho_\mathrm{PV}$ denotes the distance from the lepton track to the primary vertex and $\Delta\rho_\mathrm{PV}$ its uncertainty.

The remaining event sample is divided into two categories, based on the total number of final state lepton candidates. The search is then performed in various final state  configurations for a set of pre-determined mass hypotheses for the $\dblh$. For each mass point, the selection criteria described in Section~\ref{an_cat} are optimized using simulations, by maximizing the signal significance by means of the following significance estimator:

\[
	S_{\mathrm{cL}} = \sqrt{2(s+b) \ln(1+s/b)-2s},
\]

where $s$ is the signal expectation and $b$ is the background expectation. The estimator comes from the asymptotic expression of significance $Z=\sqrt{2\log Q}$, where $Q$ is the ratio of Poisson likelihoods $P(\text{obs} | s+b)$ and $P(\text{obs} | b)$. The estimator $S_{\mathrm{cL}}$ applies in the case of a counting experiment without systematic errors. We do not consider systematic errors at this stage as we select optimal cuts within the top 10\% of the significance across mass points and the small variations coming from systematic uncertainties do not change the optimization significantly. The $\mathrm{c}$ and $\mathrm{L}$ subscripts refer to counting experiment and likelihood, respectively. The size of the mass window is a part of the optimization procedure and is limited by the mass resolution of the signal.
\section{Analysis categories}
\label{an_cat}
The analysis is separated into categories based on the total number of light leptons and $\tauhad$ in the reconstructed events.

The decay channel with $ \mathcal{B}$$(\dblh\rightarrow\tau\tau)=100\%$ is handled separately, since the
event topology is somewhat different from the final states with prompt decays to light leptons.
In particular, the $\dblh$ reconstructed mass peak has a much larger width due to final-state
neutrinos, which affects the choice and optimization of the event selection criteria.

The final signal efficiency depends on the $\dblh$ production mechanism, decay channel and chosen mass point. 
For pair-production process and 200~\GeV $\dblh$ mass the selection efficiency varies from about $62\%$ in the $e\mu$ 
channel to $16\%$ in $\ell\tau$ channels and only $4\%$ in the $\tau\tau$ channel. Lower efficiency in decay 
channels that involve $\tau$-leptons results from the tau ID efficiency, tighter selection criteria and the 
requirement of two light leptons at the trigger level. The efficiencies slightly increase at higher mass assumptions. 
For associated production process the selection efficiencies are decreased by about a factor of two.

\subsection{\texorpdfstring{$\ell\ell\ell$ and $\ell\ell\tauhad$ final states}{Three (e|mu) and two (e|mu) tau(hadronic) final states}}
\label{3l-section}
These final states are relevant for both $\dblh$ production mechanisms. The associated production process yields three charged leptons and a neutrino. The pair-production process can contribute to this category if one of the four leptons is lost due to lepton identification inefficiency or detector acceptance.

In order to separate signal from background, a set of selection criteria is optimized for significance for various combinations
of final states and mass hypotheses. Three main categories of final states are considered: $\dblh$ decays to light leptons (ee, e$\mu$ and $\mu\mu$), $\dblh$ decays to a light-lepton and a $\tau$-lepton ($\Pe\tau$, $\mu\tau$) and $\dblh$ decay to $\tau$-leptons ($\tau\tau$). Both hadronic and  leptonic $\tau$ decays are
considered. At least two light leptons in the final state are required because of trigger considerations.

Because of the high mass of the $\dblh$, its decay products are very energetic, allowing for signal separation
through requirements on the scalar $\pt$ sum of the three leptons ($\sum\pt$) as a function of $m_\Phi$.  In addition, as a number of
important background processes contain a Z boson, events with opposite-sign same-flavor light lepton combinations are rejected if
$|m(\ell^+\ell^-)-m_{\cPZ}|$ is below a channel-dependent threshold.

A selection on the opening angle between the same-charge leptons, $\deltaphi$, is also applied.
Background processes, such as the production of a Z boson recoiling from a jet misidentified as a
lepton, yield leptons with a larger opening angle than those originating from Z decay. For the pair-production of two signal particles we expect both lepton pairs to be boosted and the opening angle to be smaller.

A loose requirement on the missing transverse energy ($\MET$), defined as the negative vectorial momentum sum of all reconstructed particle candidates, is applied in the e$\tau, \mu\tau$ and $\tau\tau$ channels in order to further reduce the background contributions, especially from Drell--Yan processes.

Finally, the mass window ($m_{\text{lower}}$, $1.1 m_\Phi$) is defined.
The lower bound, $m_{\text{lower}}$, depends on the final state. The mass windows are chosen by requiring high efficiency for signal events across a variety of final states (including $\tau$ leptonic decays, which contribute significantly in some scenarios), while keeping the analysis independent of the assumed relative branching fractions. The selection criteria used in this category are summarized in Table~\ref{tab:selection3l}.

For the 100\% branching fraction scenarios, both signal and background events are filtered based on the leptonic content. For example, when showing results for 100\% branching fraction to electrons, only events containing electrons are used. For the four benchmark points, the contributions from all possible lepton combinations are taken into account and added to the relevant distributions according to the relative branching fractions. The selection criteria of $\Pe\tau$ and $\mu\tau$ channel are used for the four benchmark points to account for various final state signatures.

\begin{figure*}[hbtp]
  \begin{center}
\includegraphics[width=0.45\textwidth]{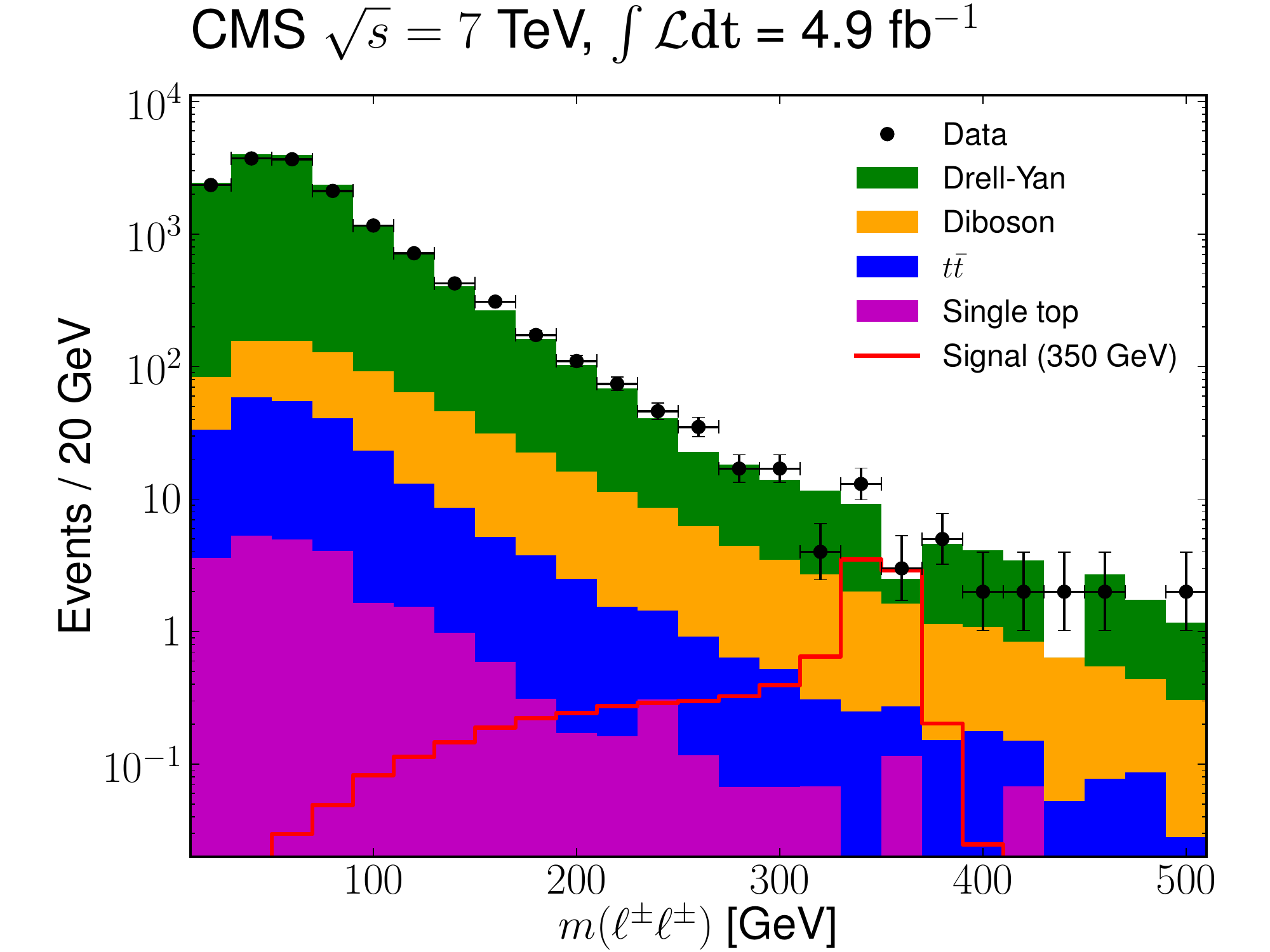}\hspace{1cm}\includegraphics[width=0.45\textwidth]{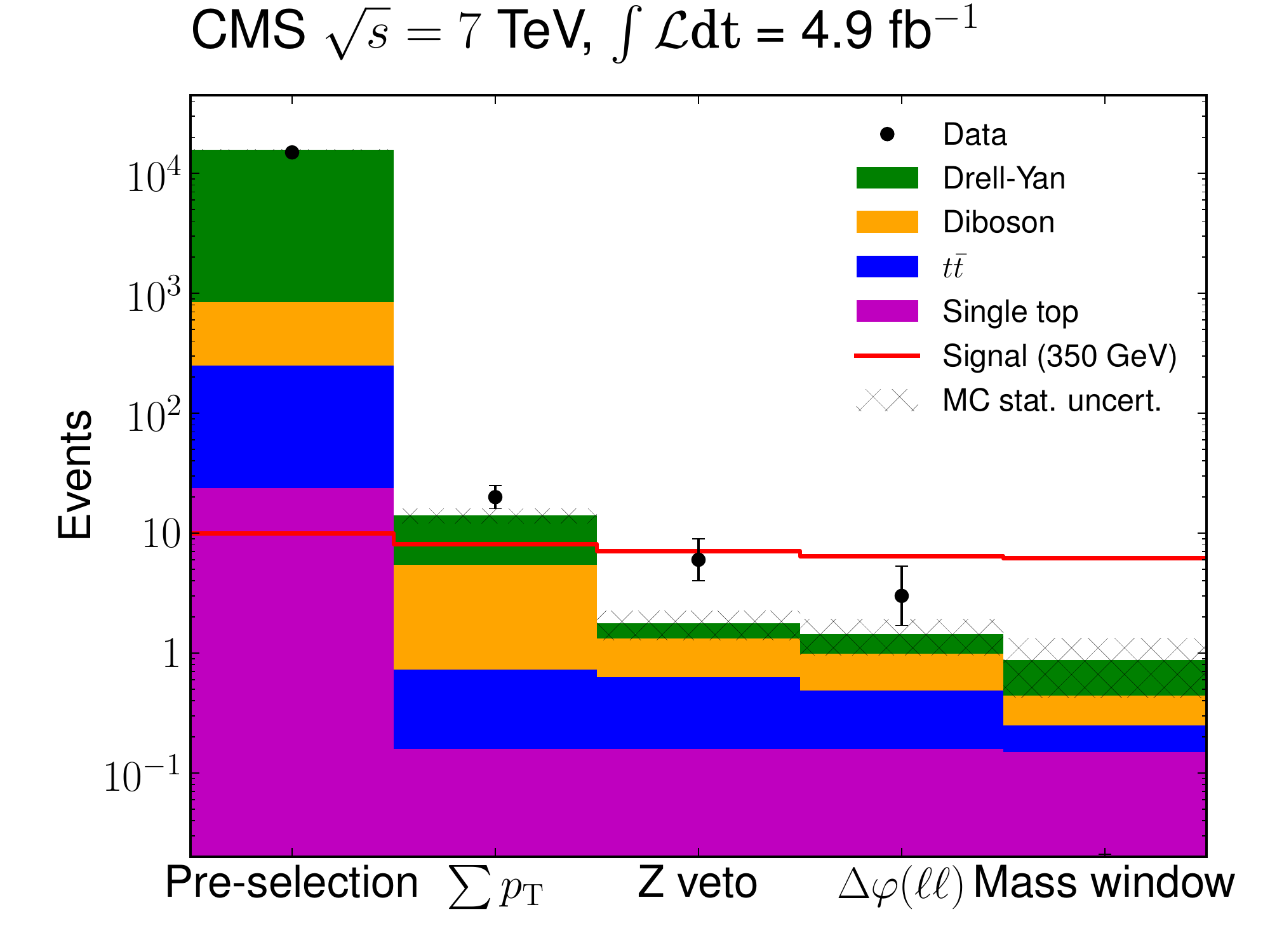}
\caption{Left: Like-charge invariant mass distribution for the $\ell\ell\ell$ and $\ell\ell\tauhad$ final state for the MC simulation and data after pre-selection. Where $\tau$ decay products are present in the final state, a visible mass is reconstructed that does not include the contribution of neutrinos. The expected distribution for a $\Phi^{++}$ with a mass of 350\GeV for the benchmark point BP4 is also shown. Right:  Event yields as a function of the applied selection criteria. $\deltaphi$ column includes both $\deltaphi$ and $\MET$ selections.}
    \label{fig:results}
  \end{center}
\end{figure*}

After the application of the selection criteria, the event yields observed in data are in reasonable agreement
with the sum of the expected contributions from backgrounds.
The mass distributions for the simulated total background and the hypothesized BP4 benchmark point signal after applying the pre-selections
are shown in Figure~\ref{fig:results}, along with the measured yields. The event yield evolution as a
function of the selections applied is also shown. For the final analysis, the background estimate is derived from data, using the methods described in Section~\ref{sec:BG}.

\begin{table*}[htbH]
\begin{center}
\topcaption{Selections applied in the three-lepton final states.}
\label{tab:selection3l}
\begin{tabular}{|c|c|c|c|}
\hline
Variable & ee, e$\mu$, $\mu\mu$ & e$\tau$, $\mu\tau$ & $\tau\tau$  \\
\hline
$\sum\pt$ & $ > 1.1 m_\Phi+60\GeV$ & $ > 0.85 m_\Phi+125\GeV$ & $> m_\Phi-10\GeV$ \\
& & &  or $>200\GeV$ \\
\hline
$|m(\ell^+\ell^-)-m_\cPZ|$ & $ > 80\GeV$ & $ > 80\GeV$ & $> 50\GeV$ \\
\hline
$\MET$ & none & $>20\GeV$ & $>40\GeV$ \\
\hline
$\deltaphi$ & $<m_\Phi/600\GeV+1.95$ & $ < m_\Phi/200\GeV+1.15$ & $<2.1$  \\
\hline
Mass window & $[0.9 m_\Phi; 1.1 m_\Phi]$ & $[m_\Phi / 2; 1.1 m_\Phi]$ & $[m_\Phi/2-20\GeV; 1.1 m_\Phi]$ \\
\hline
\end{tabular}
\end{center}
\end{table*}

\subsection{\texorpdfstring{$\ell\ell\ell\ell$, $\ell\ell\ell\tauhad$ and $\ell\ell\tauhad\tauhad$ final states}{Four (e|mu), three (e|mu) tau(hadronic), and two (e|mu) two tau(hadronic) final states}}
The requirement of a fourth lepton substantially reduces the background. The Z veto is not applied for scenarios involving only light-leptons because of low signal efficiency.

\begin{figure*}[hbtp]
  \begin{center}
    \includegraphics[width=0.45\textwidth]{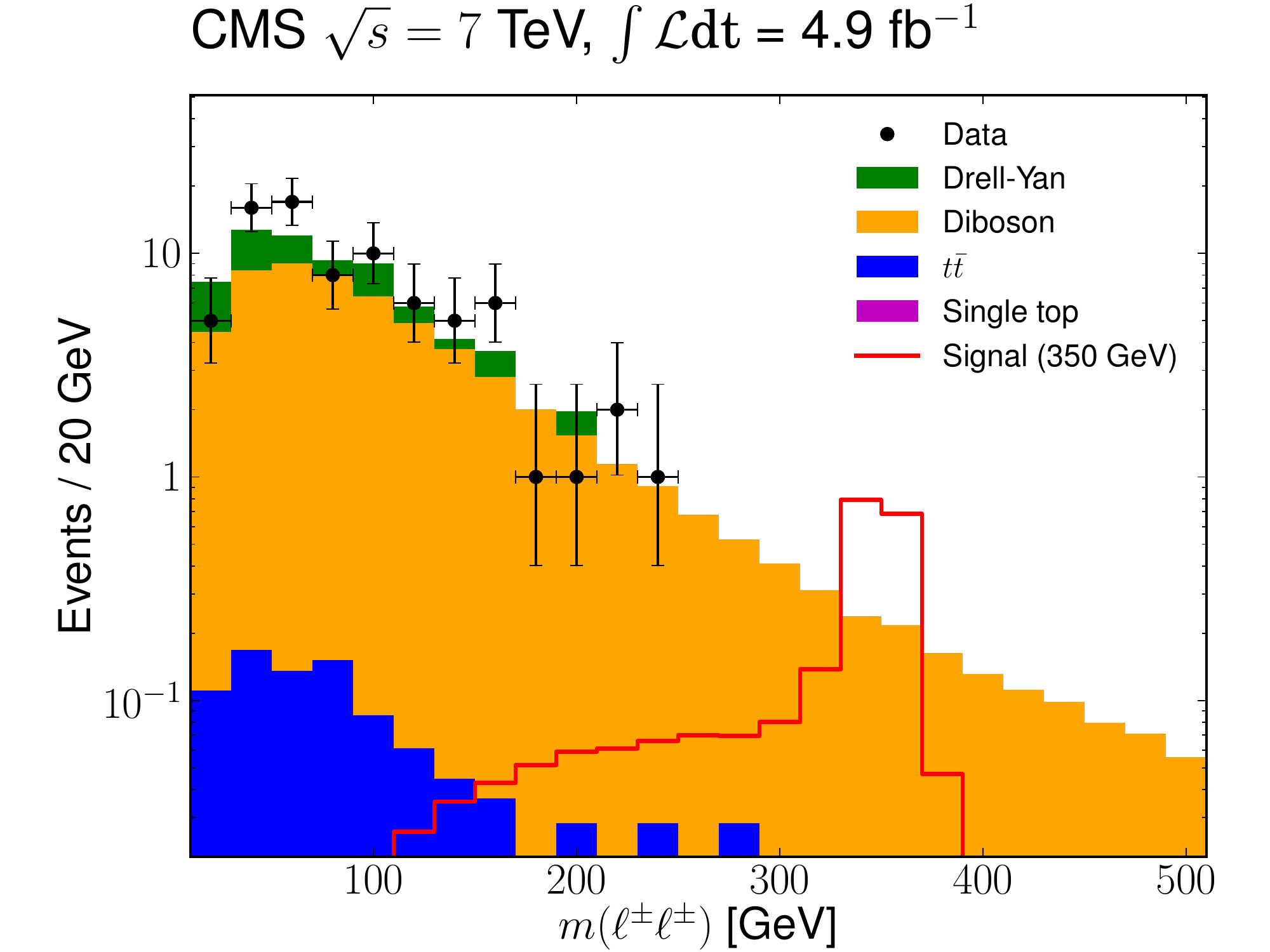}\hspace{1cm}\includegraphics[width=0.45\textwidth]{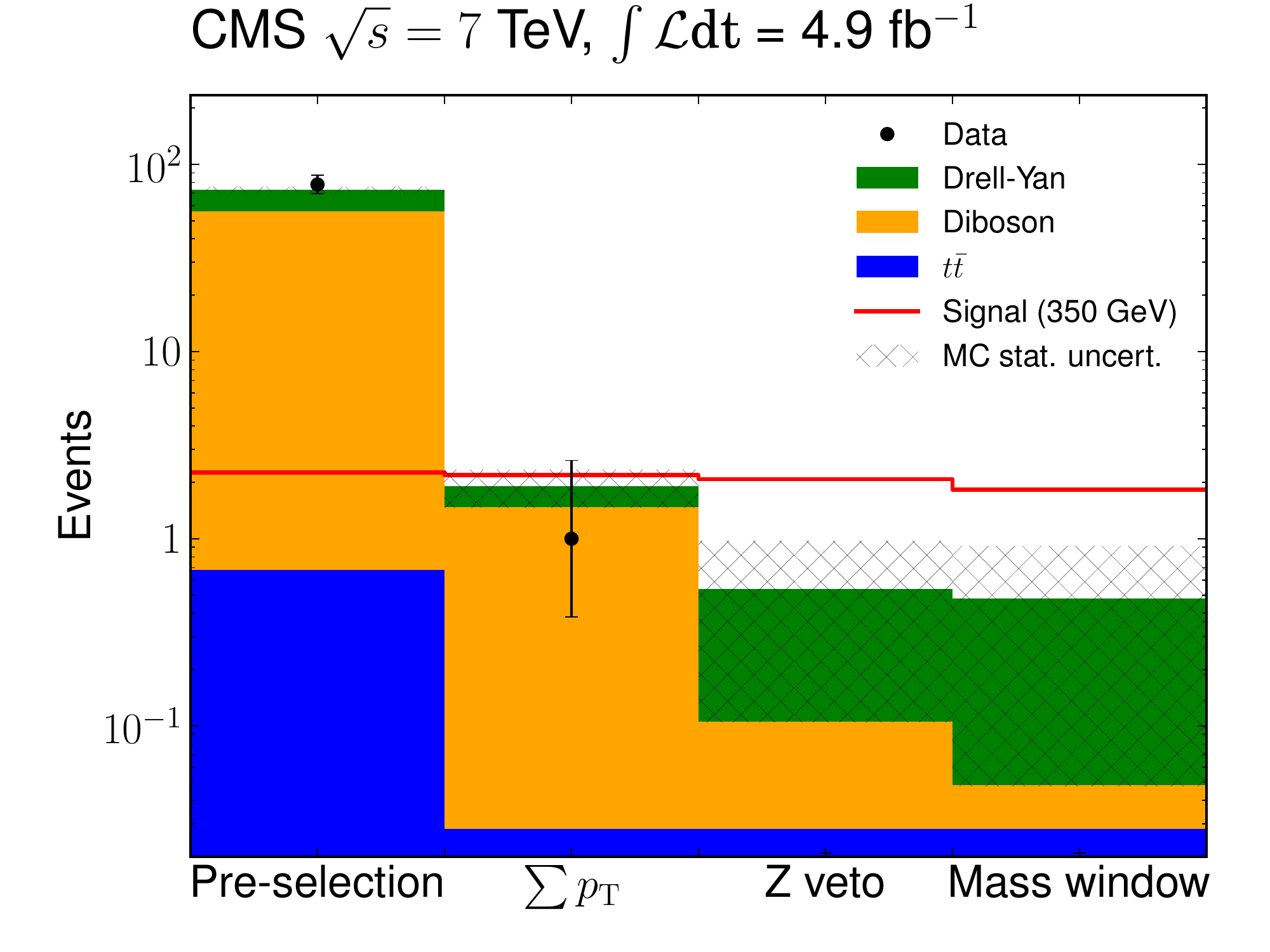}
    \caption{Left: Like-charge invariant mass distribution for the four-lepton final state for MC simulation and data after pre-selection. Where $\tau$ decay products are present in the final state, a visible mass is reconstructed that does not include the contribution of neutrinos.
The expected distribution for a $\Phi^{++}$ with a mass of 350\GeV for the benchmark point BP4 is also shown. Right: Event yields as a function of the applied selection criteria.}
    \label{fig:results4}
  \end{center}
\end{figure*}

A mass window around the doubly charged Higgs boson mass hypothesis is defined. It consists of a
two-dimensional region in the plane of $m(\ell^+\ell^+)$ vs. $m(\ell^-\ell^-)$, where $m(\ell^+\ell^+)$ and
$m(\ell^-\ell^-)$ denote the reconstructed same-sign dilepton masses. The window boundaries are the same as in Section \ref{3l-section}.
Because of the large width of the reconstructed mass peak, the mass window is not selected in the case of $ \mathcal{B}$$(\dblh\rightarrow\tau^+\tau^+)=100\%$
in order to keep the signal efficiency high. The selection criteria used in this category are summarized in Table~\ref{tab:selection4l}.
The resulting mass distributions are shown in Figure~\ref{fig:results4}.
Good agreement is seen between the event yields observed in the data and the expected background contributions.

\begin{table*}[htbH]
\begin{center}
\topcaption{Selections applied in various four-lepton final states.}
\label{tab:selection4l}
\begin{tabular}{|c|c|c|c|}
\hline
Variable & ee, e$\mu$, $\mu\mu$ & e$\tau$, $\mu\tau$ & $\tau\tau$ \\
\hline
$\sum\pt$ & $ > 0.6 m_\Phi+130\GeV$ & $>m_\Phi+100\GeV$ or $>400\GeV$ & $> 120\GeV$\\
\hline
$|m(\ell^+\ell^-)-m_{Z^0}|$ & none & $ > 10\GeV$ & $> 50\GeV$ \\
\hline
$\deltaphi$ & none & none & $ <2.5$ \\
\hline
Mass window & $[0.9 m_\Phi; 1.1 m_\Phi]$ & $[m_\Phi / 2; 1.1 m_\Phi]$ & none \\\hline
\end{tabular}
\end{center}
\end{table*}

\section{Background estimation from data}\label{sec:BG}
\subsection{Sideband method}
A sideband method is used to estimate the background contribution in the signal region. The sideband content is
determined by using same-charge di-leptons with invariant mass in the ranges $(12\GeV,$ $m_\text{lower})$ and
$(1.1 m_\Phi, 500\GeV)$ for the three-lepton final state selection. In the case of the four-lepton final state,
the sidebands comprise the $\Phi^{++}$ and $\Phi^{--}$ two-dimensional mass
plane with dilepton invariant masses between 12\GeV and 500\GeV, excluding
the candidate mass region. The upper bound of $500\GeV$ is chosen due to the
negligible expected yields for both signal and background at higher masses,
for the data sample used.

The sideband content is determined after the preselection requirements in order to ensure a reasonable number of events.
For each $\dblh$ mass hypothesis, the ratio of the event yields in the signal region to those in the sideband, $\alpha$, is estimated from the sum of all SM background MC processes:

\[
\alpha=\frac{N_\text{SR}}{N_\text{SB}},
\]

where $N_\text{SR}$ and $N_\text{SB}$ are the event yields in the signal and sideband regions respectively, estimated from simulated event samples. Modifications to this definition are made in the case of very low event counts:
\begin{itemize}
\item If $N_\text{SB} = 0$, then $\alpha=N_\text{SR}$ is assumed
\item If $N_\text{SR}$ is less than the statistical uncertainty, then the statistical uncertainty of the simulated samples is used as an estimate for the signal
region.
\end{itemize}

With an observation of $N_\text{SB}^\text{Data}$ in a sideband, the probability density function for the expected event rate is the Gamma distribution with mean $(N_\text{SB}^\text{Data}+1)$ and dispersion $\sqrt{N_\text{SB}^\text{Data}+1}$~\cite{Cousins:2008zz}. The predicted background contribution in the signal region is given by:

\[
N_\text{BGSR} = \alpha\cdot (N_\text{SB}^\text{Data}+1),
\]

with a relative uncertainty of $1/\sqrt{N_\text{SB}^\text{Data}+1}$, where $N_\text{BGSR}$
is the number of background events in the signal region estimated from the data, and $N_\text{SB}^\text{Data}$
is the total number of data events in the sidebands after applying the perselection requirements.
Where the background estimate in the signal region is smaller than the statistical uncertainty of the MC prediction,
then it is assumed that the background estimate is equal to its statistical uncertainty.

Independently of this method, control regions for major backgrounds (\ttbar, Z+jets) are defined to verify the reliability of the simulation tools in describing the data, and good agreement is found.

\subsection{ABCD method}
As a mass window is not defined for the $4\tau$ analysis, and comprises too large an area in the background region for the $3\tau$
analysis with $m_{\dblh}<200\GeV$, the sideband method cannot be used for these modes. Instead, we use the
'ABCD method', which estimates the number of background events after the final selection (signal region A) by
extrapolating the event yields in three sidebands (B, C and D).
The signal region and three sidebands are defined using a set of two observables $x$ and $y$, that define four exclusive regions in the
parameter space. The requirement of negligible correlation between $x$ and $y$ ensures that the probability density function of the background
can be factorized as $\rho(x,y) = f(x)g(y)$. It can be shown that the expectation values
of the event yields in the four regions fulfill the relation $\lambda_A/\lambda_B = \lambda_D/\lambda_C$.
The quantities $\lambda_X$ are the parameters of the Poisson distribution, which for one measurement correspond to the event counts $N_X$.
The estimated number of background events in the signal region is then given by

\[
N_A = N_B\cdot\frac{N_D}{N_C}.
\]

The variables $\RelIso$ and $|m(\ell^{+}\ell^{-})-m_\mathrm{Z}|$ for the $3\tau$ analysis and $\RelIso$ and $\sum \pt$ for
the $4\tau$ analysis are chosen based on their low correlation and the available amount of data in the sidebands.
High values of RelIso populate the sidebands with background events, where jets have been misidentified as leptons.
Failing the $|m(\ell^{+}\ell^{-})-m_{Z^0}|>50$\GeV requirement gives mainly background contributions
from the Drell-Yan and di-boson processes, whereas low values of $\sum \pt$ can probe various background processes
that possibly contain genuine leptons, but do not belong to the signal phase space.

The estimated number of background events agrees well with both the prediction from simulation and the number of data events observed in
the signal region.

\section{Systematic uncertainties}
The impact on the selection efficiency of the uncertainties related to the electron and muon identification
and isolation algorithms, and the relevant mis-identification rates, detailed
in~\cite{EGM-10-004,MUO-10-002,Chatrchyan:2011nx,TAU-11-001,HIG-11-025}, are estimated to
be less than $2\%$ using a standard `tag-and-probe' method \cite{CMS:2011aa} that relies upon $\cPZ \rightarrow \ell^+\ell^-$ decays to provide an unbiased and high-purity sample of leptons. A `tag' lepton is required to satisfy stringent criteria on reconstruction, identification, and isolation, while a `probe' lepton is used to measure the efficiency of a particular selection by using the Z mass constraint. 
The $2\%$ uncertainty that is assigned to lepton identification comprises also the charge misidentification uncertainty.
The ratio of the overall efficiencies as measured in data and simulated events is used as a correction
factor in the bins of $\pt$ and $\eta$ for the efficiency determined through simulation, and is propagated to the final result.

The $\tauhad$ reconstruction and identification efficiency via the HPS algorithm is also derived from data and simulations, using the tag-and-probe method with $\cPZ\rightarrow\tau^{+}(\rightarrow\mu^++\Pagngm+\Pgngt)\tau^{-}(\rightarrow \tauhad +\Pgngt)$ events~\cite{TAU-11-001}. The uncertainty of the measured efficiency of the $\tauhad$ algorithms is 6\%~\cite{TAU-11-001}.
Estimation of the $\tauhad$ energy-scale uncertainty is also performed with data in the
$Z \rightarrow \tau \tau \rightarrow \mu + \tauhad$ final state, and is found to be less than 3\%. The $\tauhad$ charge misidentification rate is measured to be less than 3\%.

The theoretical uncertainty in the signal cross section, which has been calculated at NLO, is about 10--15\%, and arises because of its sensitivity to the renormalization scale and the parton distribution functions (PDF)~\cite{Muhlleitner:2003me}.

The ratio $\alpha$ used to estimate the background contribution in the signal region is affected by two main uncertainties.
The first is based on the uncertainty of the ratio of the simulated event yields in the sideband and the signal regions,
and is related to the size of the kinematic region defined by the selection criteria. This uncertainty is dominated by the
PDF and renormalization scale, in addition to the lepton energy scales. The combined uncertainty is 5\%~\cite{EWK-10-013}. The other component comes from the statistical uncertainty of the small event content of the sidebands. This uncertainty is as high as 100\% if no events are observed in data.
The luminosity uncertainty is estimated to be 2.2\%~\cite{CMS-PAS-SMP-12-008}.

The systematic uncertainties are summarized in Table~\ref{tab:systematics}.
The first eight rows in the table concern the signal and the final two rows the background processes.
Correlations of systematic uncertainties within a given decay mode and between different modes are taken into account in the limit calculations.

\begin{table*}[htbp!]
\begin{center}
\topcaption{Impact of systematic uncertainties.}
\label{tab:systematics}
\begin{tabular}{|c|c|}
\hline
Lepton (e or $\mu$) ID and isolation & 2\% \\
$\tauhad$ ID and isolation & 6\%\\
$\tauhad$ energy scale & 3\%\\
$\tauhad$ misidentification rate & 3\%\\
Trigger and primary vertex finding & 1.5\% \\
Signal cross section & 10\% \\
Luminosity & 2.2\% \\
Statistical uncertainty of signal samples & 1-7\%\\
Ratio used in background estimation & 5-100\% \\
Statistical uncertainty of observed data events in sideband & 10-100\% \\
\hline
\end{tabular}
\end{center}
\end{table*}
\section{Results and statistical interpretation}
\begin{figure*}[hbtp]
  \begin{center}
    \includegraphics[width=\textwidth]{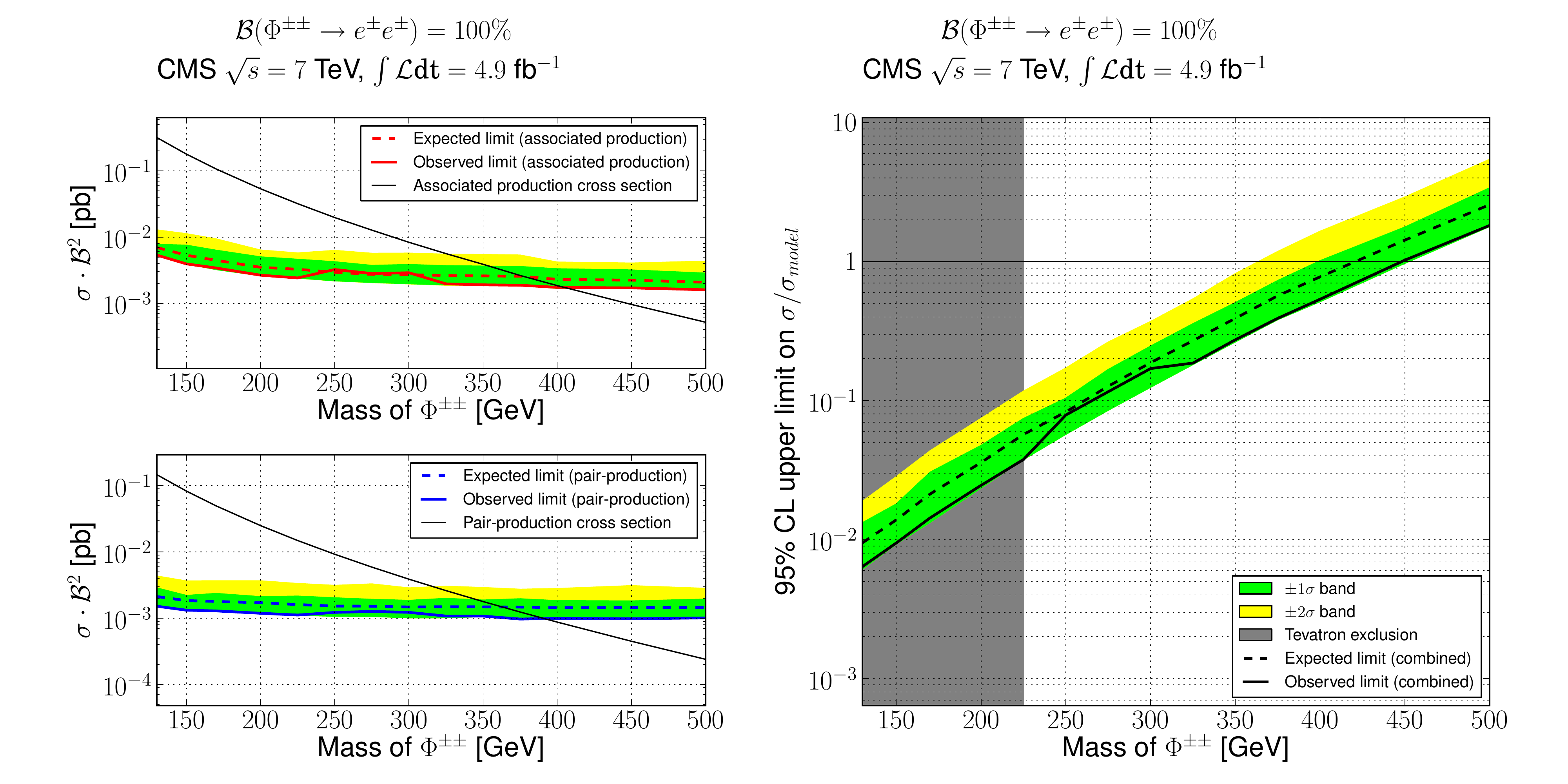}
    \caption{Lower bound on $\dblh$ mass at 95\% CL for ${\cal B}(\dblh\rightarrow e^+e^+)=100\%$.}
    \label{fig:CLs}
  \end{center}
\end{figure*}

\begin{figure*}[hbtp!]
  \begin{center}
    \includegraphics[width=\textwidth]{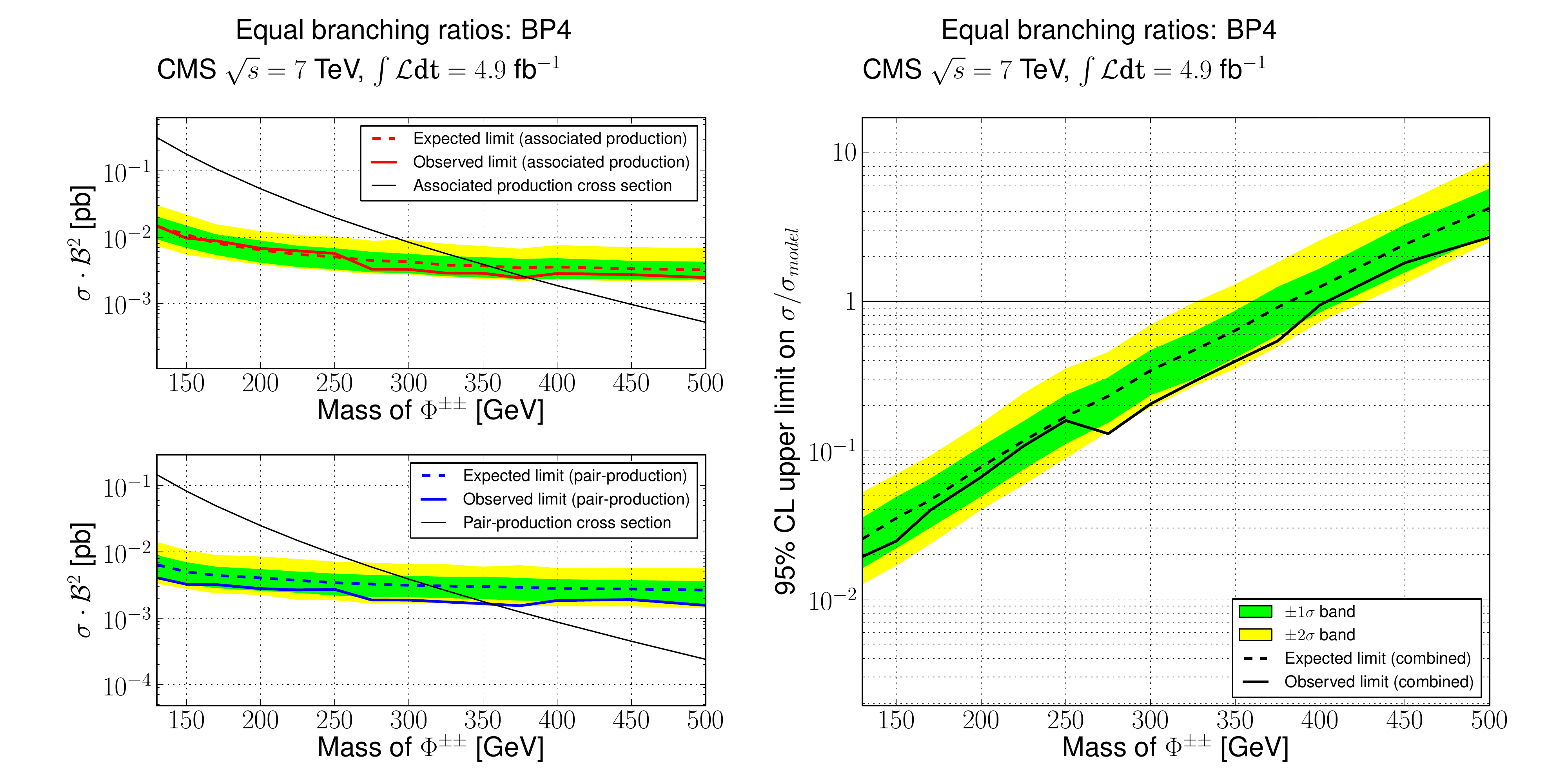}
    \caption{Lower bound on $\dblh$ mass at 95\% CL for BP4. On the left hand plots the ${\cal B}^2$ means ${\cal B}(\dblh\rightarrow\ell_\alpha^+\ell_\beta^+) {\cal B}(\dblh\rightarrow\ell_\gamma^+\ell_\delta^+)$ summed over all possible flavor combinations.}
    \label{fig:CLs2}
  \end{center}
\end{figure*}

The data and the estimated background contributions are found to be in reasonable agreement for all final states.
Only a few events are observed with invariant masses above 200\GeV, consistent with SM background expectations.
The dataset is used to derive limits on the doubly-charged Higgs mass in all decay channels.
A $\mathrm{CL_{S}}$ method \cite{LEP-CLs} is used to calculate an upper limit for the $\dblh$ cross section at the $95\%$ confidence level (CL),
which includes the systematic uncertainties summarized in Table~\ref{tab:systematics}. As the
systematic uncertainties are different for each final state, the signal and background yields
are separated into five orthogonal categories, based on the number of light leptons and $\tau$-leptons.
As an example, event yields in four mass points for BP4 can be found in Table~\ref{tab:bgest-BP4}. A full list of mass points 
considered for the limit calculation is given in the end of Section~\ref{sec:MC}.
When setting limits on `muon and electron only' channels, we only distinguish the cases of three and
four leptons with no $\tauhad$ involved. The limits are interpolated linearly.
The results of the exclusion limit calculations are reported in Figures~\ref{fig:CLs}-\ref{fig:CLs2}, and summarized in Table~\ref{tab:worldlimits}.

The cross section limits significantly improve on previously published lower bounds on the $\dblh$ mass.
New limits are also set on the four benchmark points, probing a large region of the parameter
space of type II seesaw models.

    \begin{table*}[p]
    \begin{center}
    \scriptsize{
    \topcaption{Background estimation from simulation and data, observed number of events, and signal yields for BP4.}
    \label{tab:bgest-BP4}
    \begin{tabular}{|c|c|clc|c|c|c|}
    \hline
    Mass & Final state & MC estimate & Estimate from data & Observed events& Pair-production & Associate production \\
\hline
200\GeV & $\ell\ell\ell$ & $0.99\pm0.43$ & $1.32\pm0.64\pm0.02$ & $2$ & $9.35\pm0.07$ & $33.17\pm0.15$ \\
200\GeV & $\ell\ell\tauhad$ & $0.52\pm0.07$ & $0.50\pm0.10\pm0.01$ & $1$ & $3.05\pm0.04$ & $8.02\pm0.08$ \\
200\GeV & $\ell\ell\ell\ell$ & $0.05\pm0.02$ & $0.07\pm0.04\pm0.01$ & $0$ & $17.25\pm0.07$ & $0.01\pm0.01$ \\
200\GeV & $\ell\ell\ell\tauhad$ & $0.03\pm0.02$ & $0.02\pm0.02\pm0.01$ & $0$ & $4.55\pm0.05$ & $0.04\pm0.01$ \\
200\GeV & $\ell\ell\tauhad\tauhad$ & $0.03\pm0.02$ & $0.02\pm0.02\pm0.01$ & $0$ & $0.57\pm0.02$ & $0.0\pm0.0$ \\
\hline
300\GeV & $\ell\ell\ell$ & $0.22\pm0.03$ & $0.30\pm0.06\pm0.01$ & $0$ & $2.06\pm0.02$ & $7.07\pm0.04$ \\
300\GeV & $\ell\ell\tauhad$ & $0.12\pm0.04$ & $0.12\pm0.04\pm0.01$ & $0$ & $0.62\pm0.01$ & $1.52\pm0.02$ \\
300\GeV & $\ell\ell\ell\ell$ & $0.03\pm0.02$ & $0.04\pm0.03\pm0.01$ & $0$ & $3.06\pm0.02$ & $0.0\pm0.0$ \\
300\GeV & $\ell\ell\ell\tauhad$ & $0.03\pm0.02$ & $0.02\pm0.02\pm0.01$ & $0$ & $0.78\pm0.01$ & $0.0\pm0.0$ \\
300\GeV & $\ell\ell\tauhad\tauhad$ & $0.03\pm0.02$ & $0.02\pm0.02\pm0.01$ & $0$ & $0.10\pm0.01$ & $0.0\pm0.0$ \\
\hline
400\GeV & $\ell\ell\ell$ & $0.19\pm0.04$ & $0.26\pm0.07\pm0.01$ & $1$ & $0.60\pm0.01$ & $1.94\pm0.01$ \\
400\GeV & $\ell\ell\tauhad$ & $0.06\pm0.02$ & $0.06\pm0.03\pm0.01$ & $0$ & $0.17\pm0.01$ & $0.4\pm0.01$ \\
400\GeV & $\ell\ell\ell\ell$ & $0.03\pm0.02$ & $0.04\pm0.03\pm0.01$ & $0$ & $0.70\pm0.01$ & $0.0\pm0.0$ \\
400\GeV & $\ell\ell\ell\tauhad$ & $0.03\pm0.02$ & $0.02\pm0.02\pm0.01$ & $0$ & $0.18\pm0.01$ & $0.0\pm0.0$ \\
400\GeV & $\ell\ell\tauhad\tauhad$ & $0.03\pm0.02$ & $0.02\pm0.02\pm0.01$ & $0$ & $0.02\pm0.01$ & $0.0\pm0.0$ \\
\hline
450\GeV & $\ell\ell\ell$ & $0.14\pm0.04$ & $0.19\pm0.06\pm0.03$ & $1$ & $0.32\pm0.01$ & $1.04\pm0.01$ \\
450\GeV & $\ell\ell\tauhad$ & $0.04\pm0.02$ & $0.04\pm0.03\pm0.00$ & $0$ & $0.08\pm0.01$ & $0.21\pm0.01$ \\
450\GeV & $\ell\ell\ell\ell$ & $0.03\pm0.02$ & $0.04\pm0.03\pm0.01$ & $0$ & $0.36\pm0.01$ & $0.0\pm0.0$ \\
450\GeV & $\ell\ell\ell\tauhad$ & $0.03\pm0.02$ & $0.02\pm0.02\pm0.01$ & $0$ & $0.09\pm0.01$ & $0.0\pm0.0$ \\
450\GeV & $\ell\ell\tauhad\tauhad$ & $0.03\pm0.02$ & $0.02\pm0.02\pm0.01$ & $0$ & $0.01\pm0.0$ & $0.0\pm0.0$ \\
    \hline
    \end{tabular}
    }
    \end{center}
    \end{table*}

\begin{table*}[htbH]
\begin{center}
\topcaption{Summary of the 95\% CL exclusion limits.}
\label{tab:worldlimits}
\footnotesize{\begin{tabular}{|c|c|c|}
\hline
Benchmark point &  Combined 95\% CL limit [\GeVns{}] & 95\% CL limit\\
 & & for pair production only [\GeVns{}]\\
\hline
$\mathcal{B}$$(\dblh\rightarrow \Pep \Pep) = 100\%$ & \eeres &  \eerespp \\
$\mathcal{B}$$(\dblh\rightarrow \Pep \mu^{+}) = 100\%$ & \emres & \emrespp  \\
$\mathcal{B}$$(\dblh\rightarrow \Pep \tau^{+}) = 100\%$ & \etres & \etrespp \\
$\mathcal{B}$$(\dblh\rightarrow \mu^{+} \mu^{+}) = 100\%$ & \mmres & \mmrespp \\
$\mathcal{B}$$(\dblh\rightarrow \mu^{+} \tau^{+}) = 100\%$ & \mtres & \mtrespp \\
$\mathcal{B}$$(\dblh\rightarrow \tau^{+} \tau^{+}) = 100\%$ & \ttres & \ttrespp \\
BP1 & \bponeres & \bponerespp \\
BP2 & \bptwores & \bptworespp \\
BP3 & \bpthreeres & \bpthreerespp \\
BP4 & \bpfourres & \bpfourrespp \\
\hline
\end{tabular}
}\end{center}
\end{table*}

\section{Summary}
A search for the doubly-charged Higgs boson $\dblh$ has been conducted using a data sample corresponding to an
integrated luminosity of \lumitot collected by the CMS experiment at a center-of-mass energy of 7\TeV. No evidence
for the existence of the $\dblh$ has been found. Lower bounds on the $\dblh$ mass are established between 204 and
459\GeV in the 100\% branching fraction scenarios, and between 383 and 408\GeV for four benchmark points of the type II seesaw model,
providing significantly more stringent constraints than previously published limits.

\begin{figure*}[hbtp]
  \begin{center}
    \includegraphics[width=\textwidth]{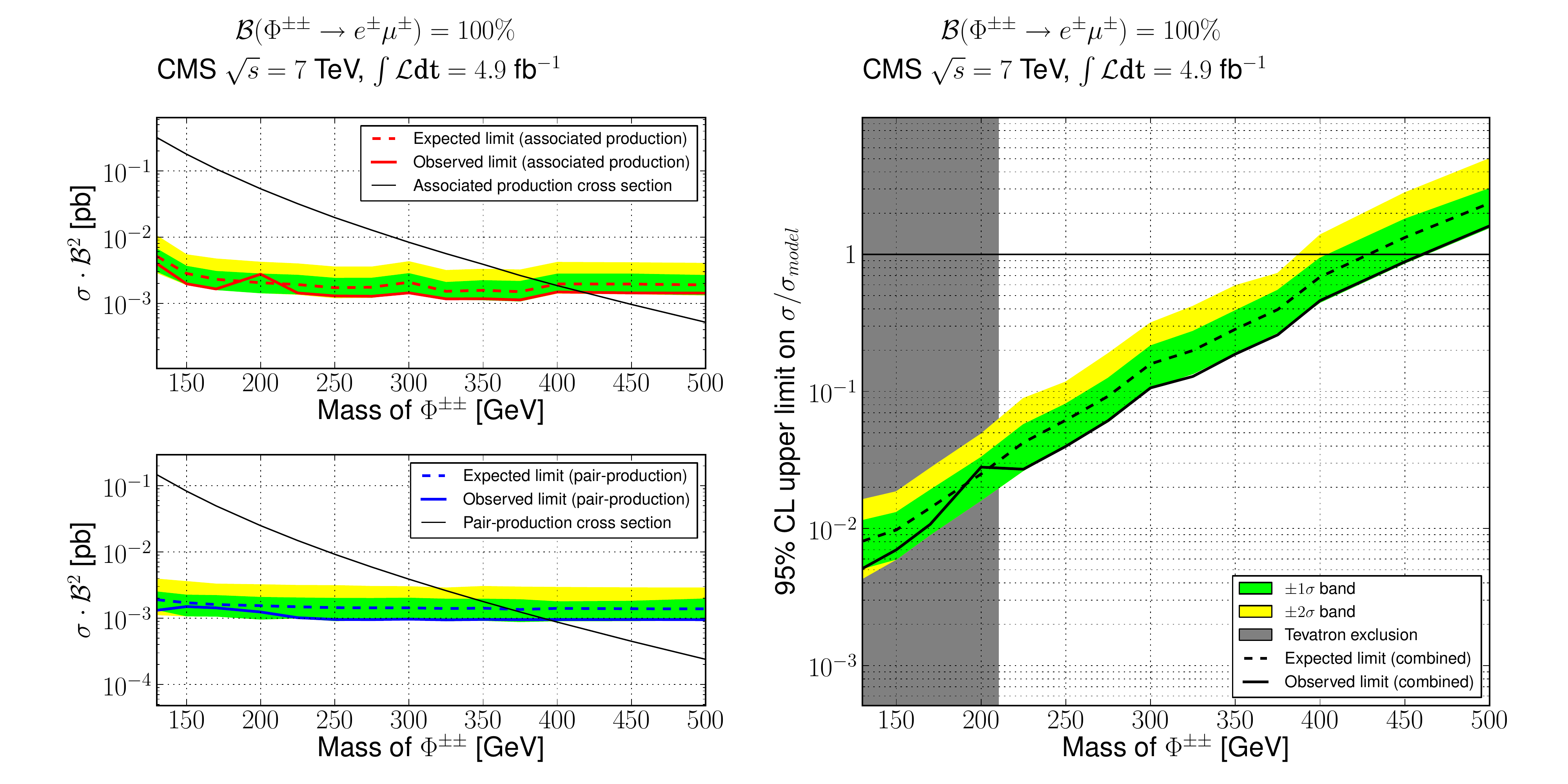}
    \caption{Lower bound on $\dblh$ mass at 95\% CL for ${\cal B}(\dblh\rightarrow e^+\mu^+)=100\%$.}
  \end{center}
\end{figure*}

\begin{figure*}[hbtp]
  \begin{center}
    \includegraphics[width=\textwidth]{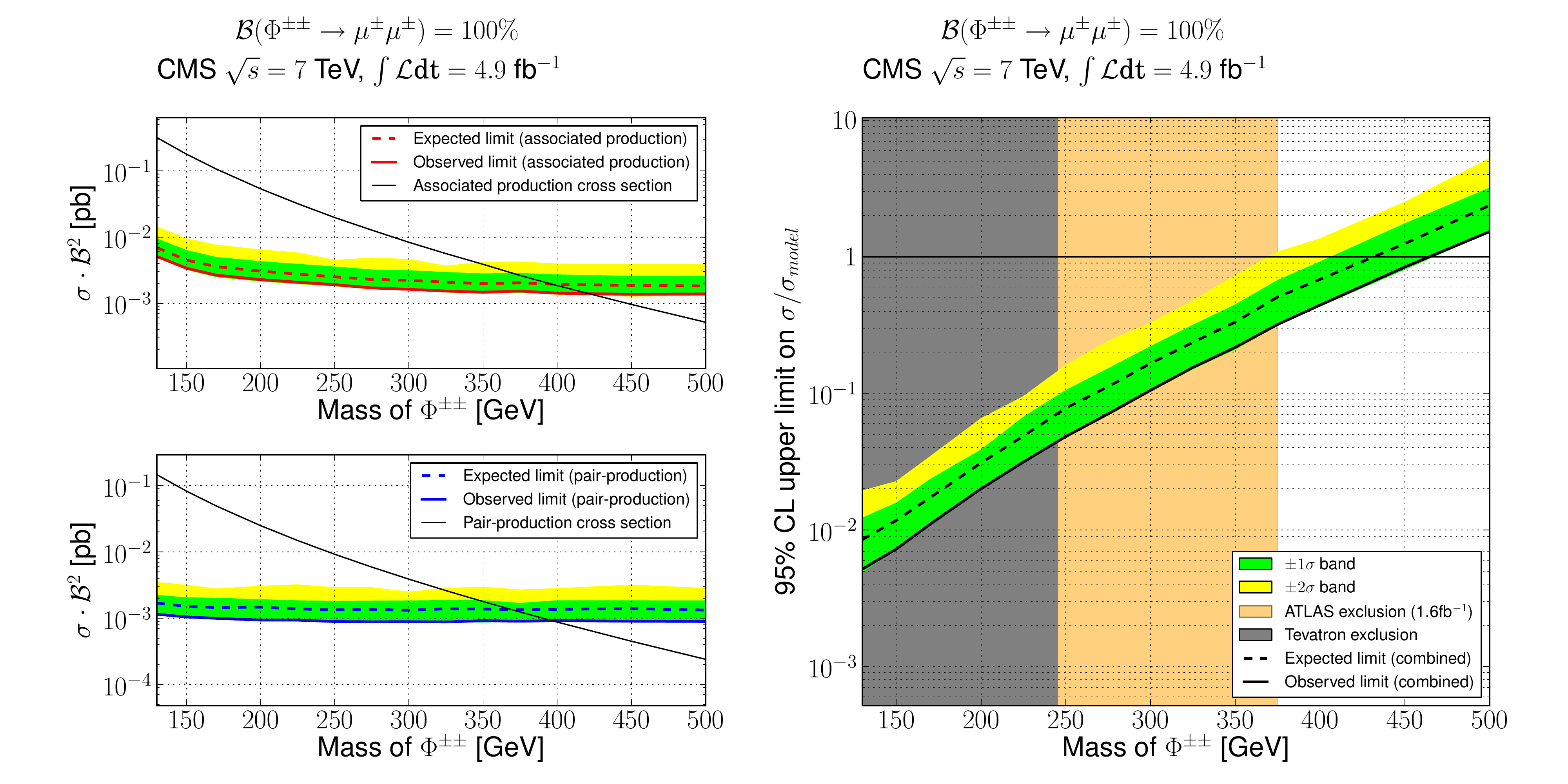}
    \caption{Lower bound on $\dblh$ mass at 95\% CL for ${\cal B}(\dblh\rightarrow\mu^+\mu^+)=100\%$.}
  \end{center}
\end{figure*}

\begin{figure*}[hbtp]
  \begin{center}
    \includegraphics[width=\textwidth]{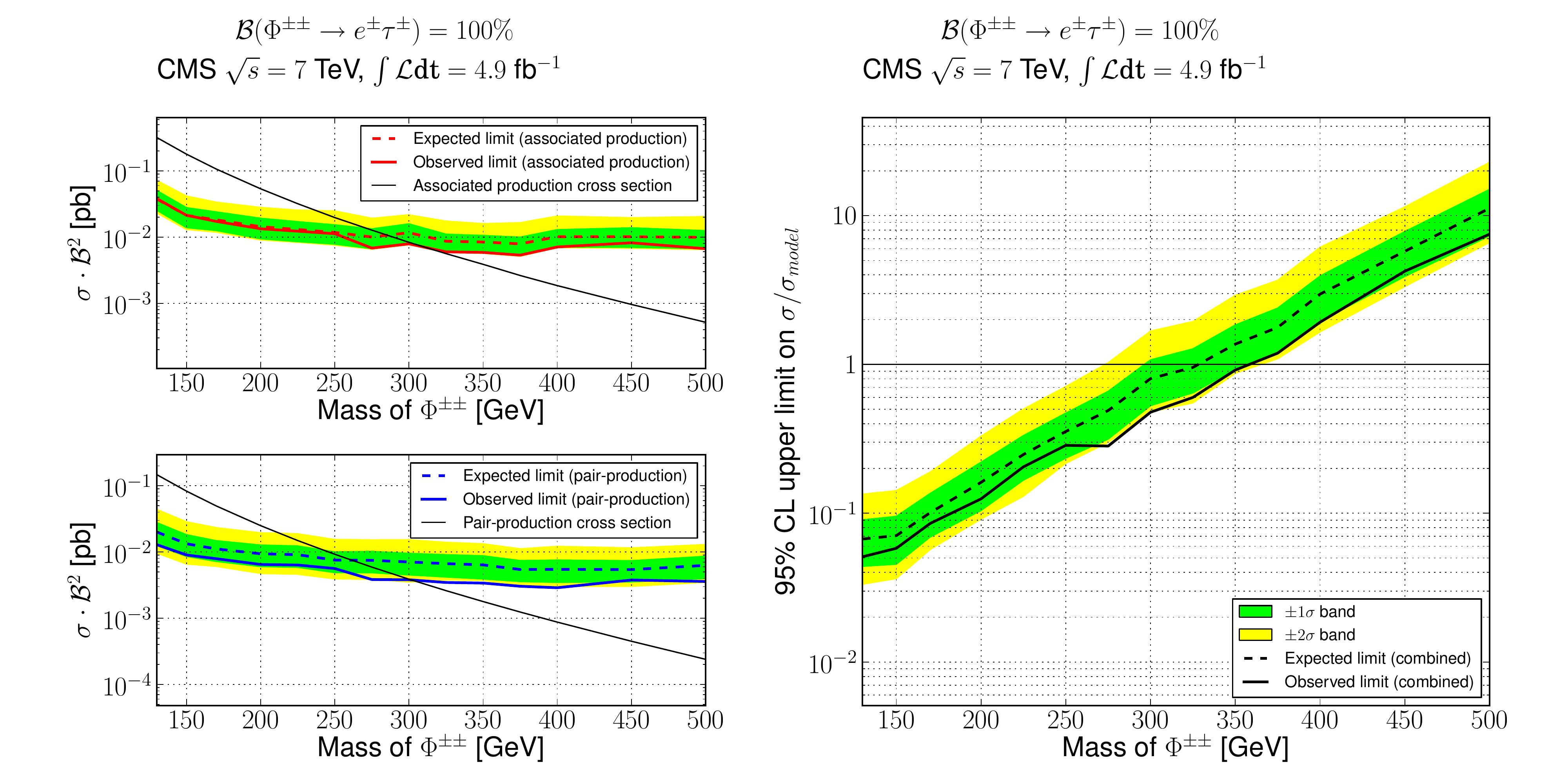}
    \caption{Lower bound on $\dblh$ mass at 95\% CL for ${\cal B}(\dblh\rightarrow e^+\tau^+)=100\%$.}
  \end{center}
\end{figure*}

\begin{figure*}[hbtp]
  \begin{center}
    \includegraphics[width=\textwidth]{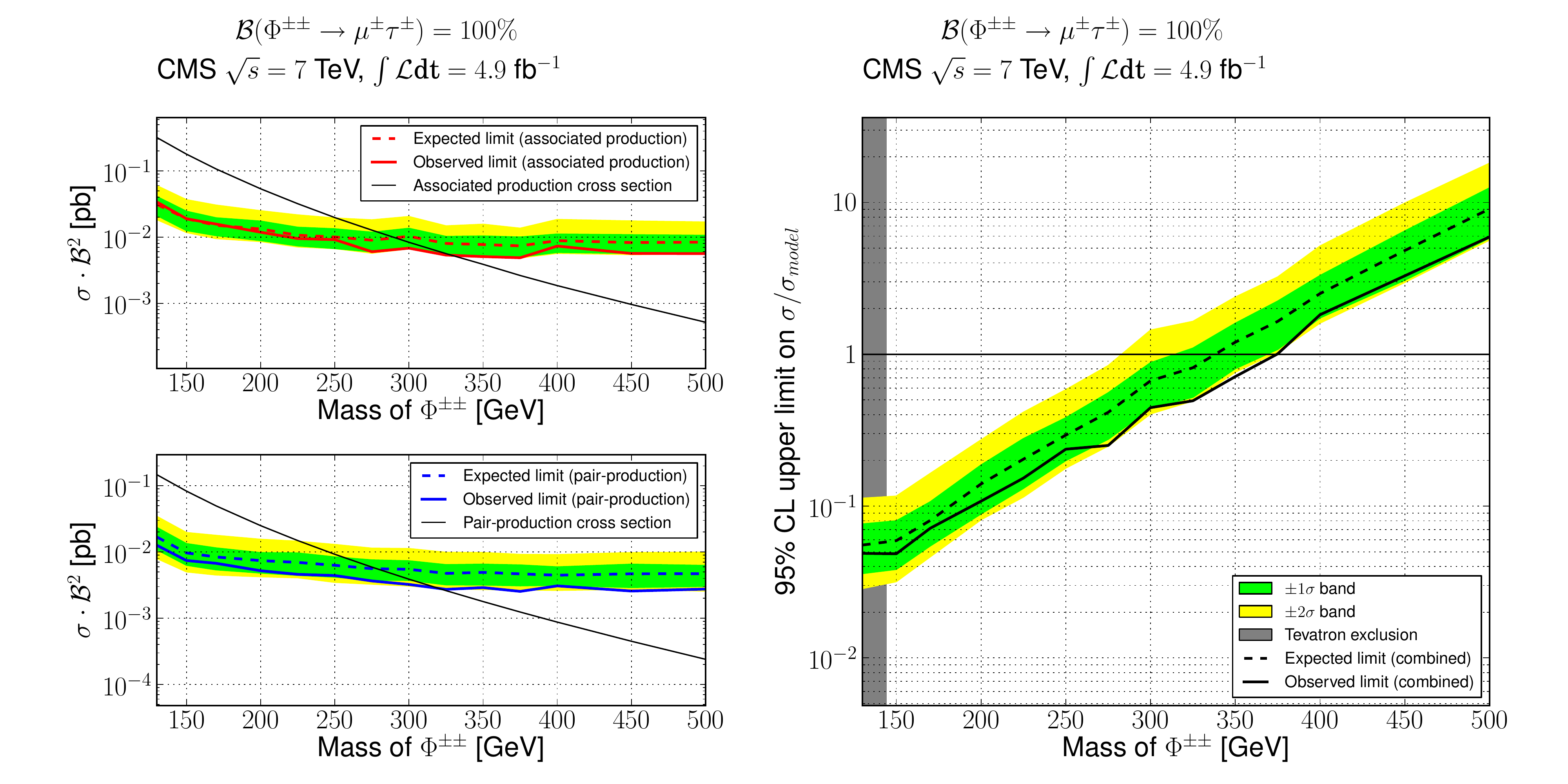}
    \caption{Lower bound on $\dblh$ mass at 95\% CL for ${\cal B}(\dblh\rightarrow\mu^+\tau^+)=100\%$.}
  \end{center}
\end{figure*}

\begin{figure*}[hbtp]
  \begin{center}
    \includegraphics[width=\textwidth]{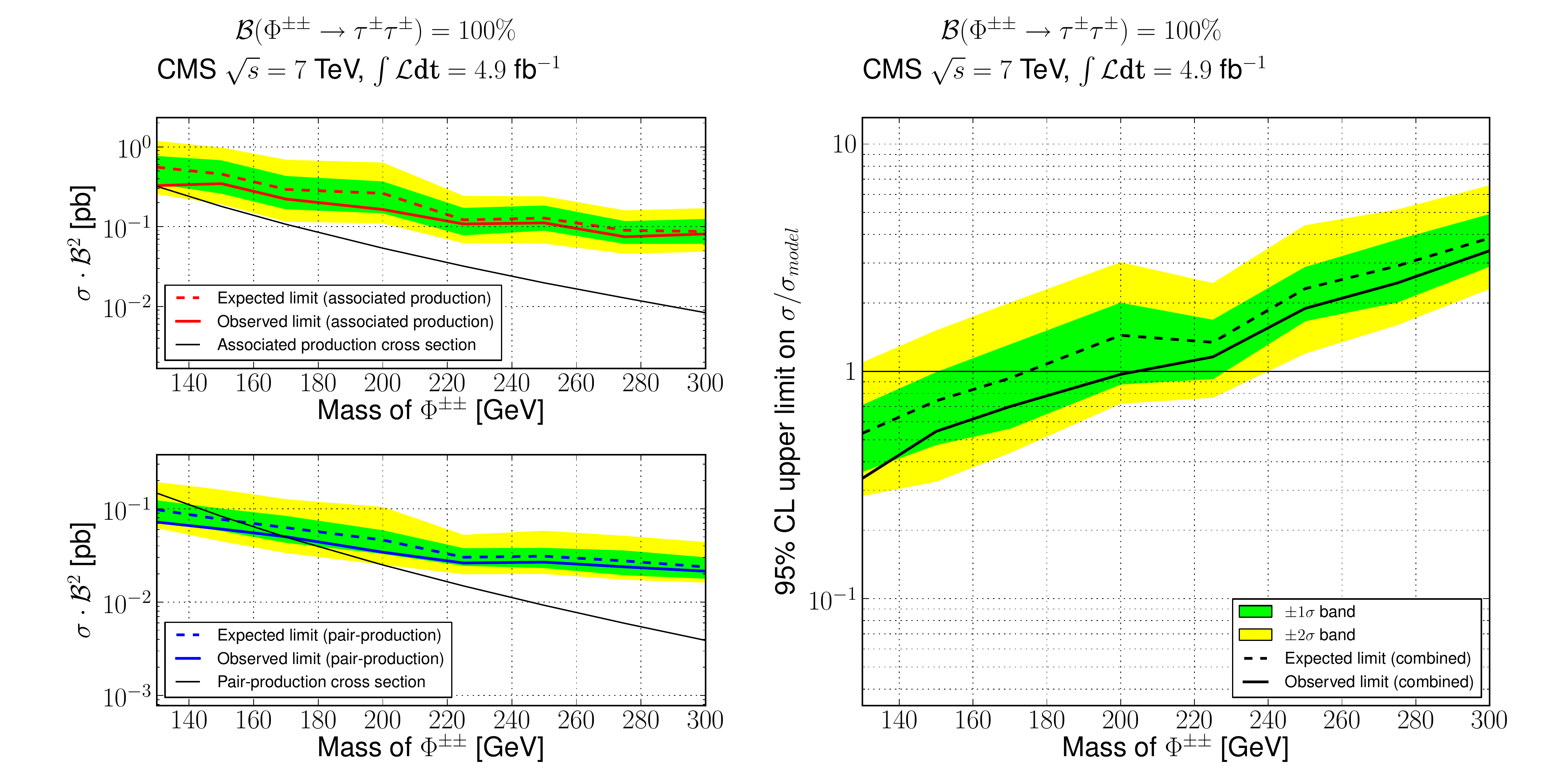}
    \caption{Lower bound on $\dblh$ mass at 95\% CL for ${\cal B}(\dblh\rightarrow\tau^+\tau^+)=100\%$.}
  \end{center}
\end{figure*}

\begin{figure*}[hbtp]
  \begin{center}
    \includegraphics[width=\textwidth]{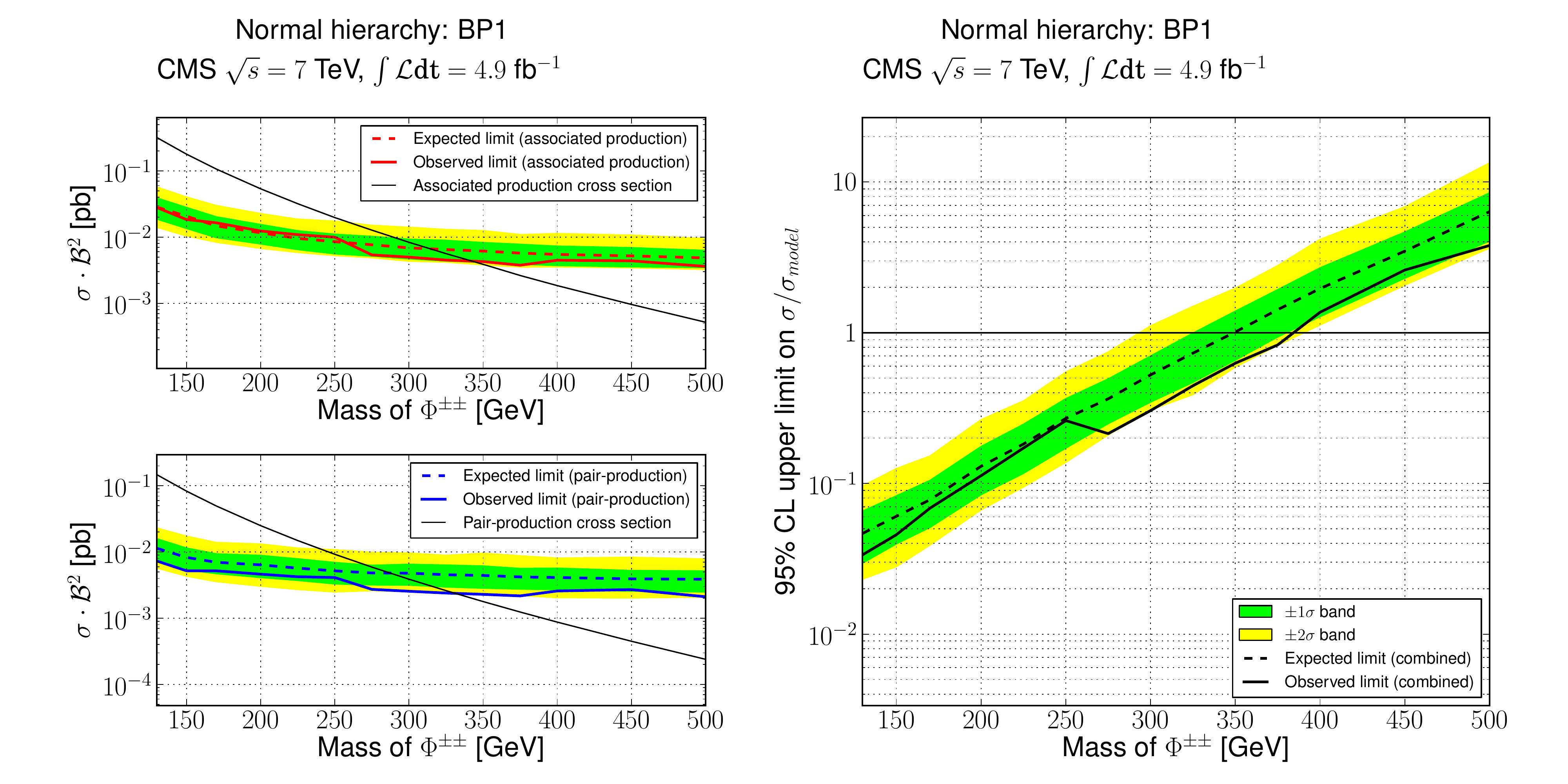}
    \caption{Lower bound on $\dblh$ mass at 95\% CL for BP1. On the left hand plots the ${\cal B}^2$ means ${\cal B}(\dblh\rightarrow\ell_\alpha^+\ell_\beta^+) {\cal B}(\dblh\rightarrow\ell_\gamma^+\ell_\delta^+)$ summed over all possible flavor combinations.}
  \end{center}
\end{figure*}

\begin{figure*}[hbtp]
  \begin{center}
    \includegraphics[width=\textwidth]{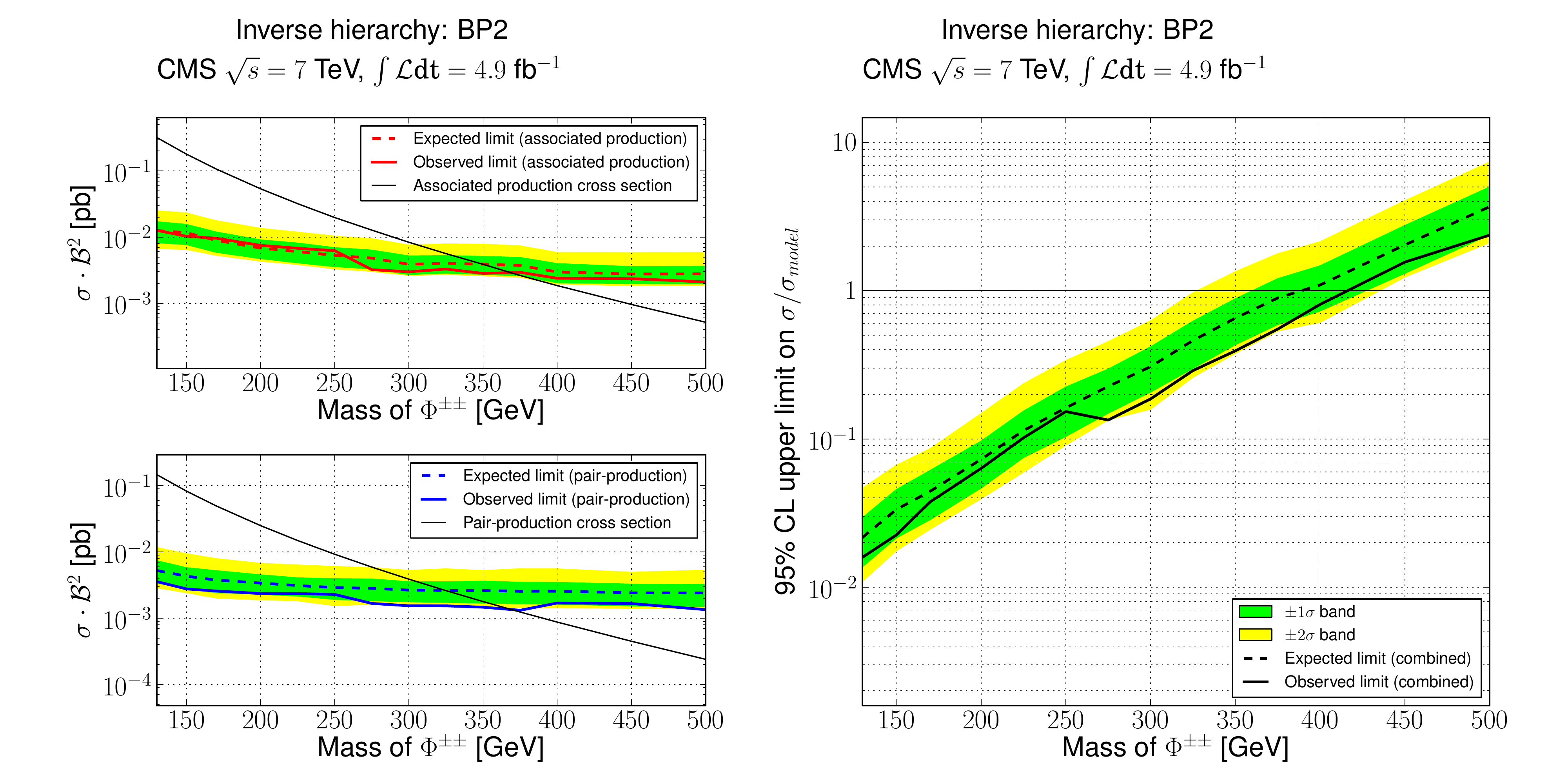}
    \caption{Lower bound on $\dblh$ mass at 95\% CL for BP2. On the left hand plots the ${\cal B}^2$ means ${\cal B}(\dblh\rightarrow\ell_\alpha^+\ell_\beta^+) {\cal B}(\dblh\rightarrow\ell_\gamma^+\ell_\delta^+)$ summed over all possible flavor combinations.}
  \end{center}
\end{figure*}

\begin{figure*}[hbtp]
  \begin{center}
    \includegraphics[width=\textwidth]{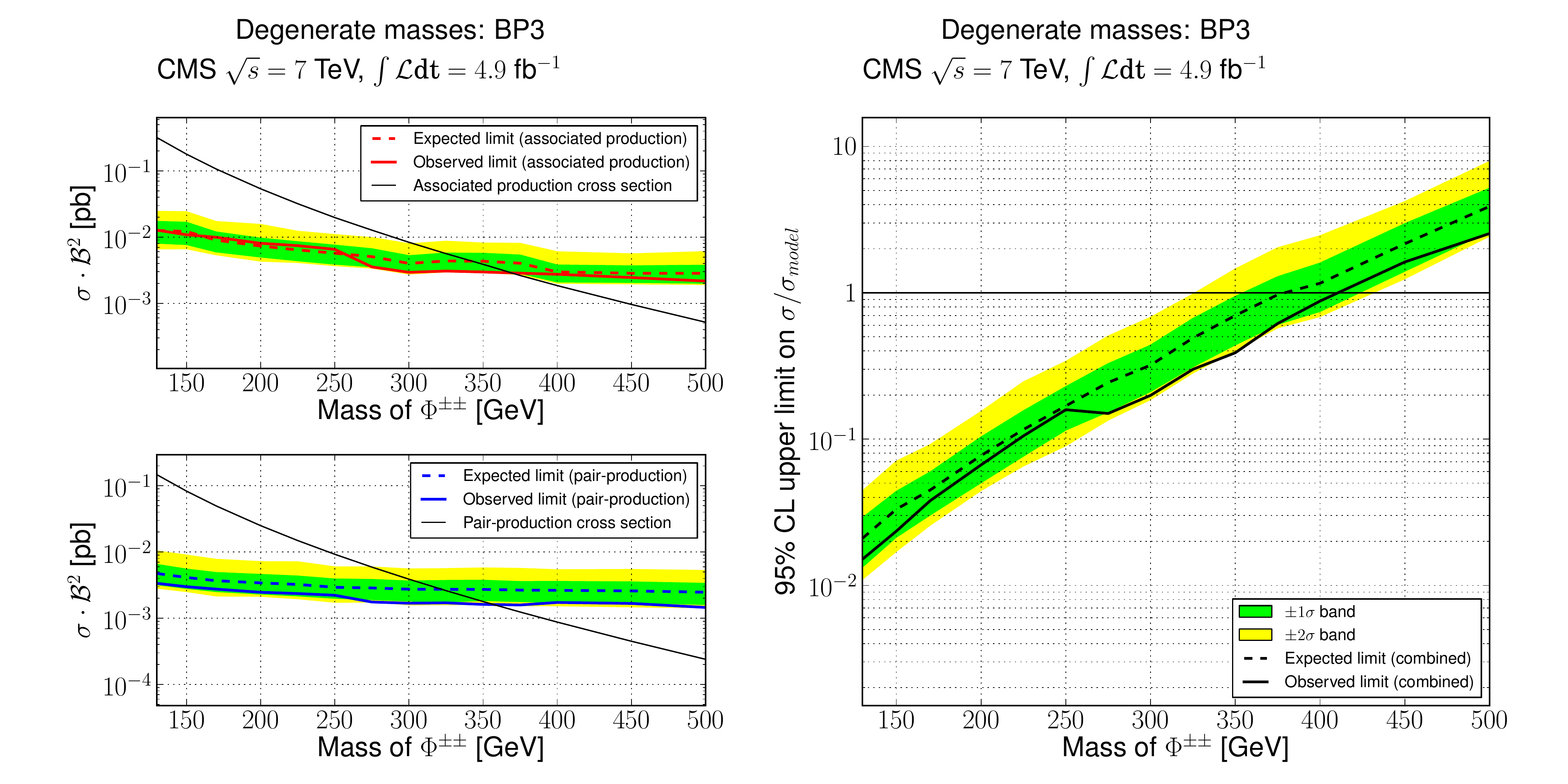}
    \caption{Lower bound on $\dblh$ mass at 95\% CL for BP3. On the left hand plots the ${\cal B}^2$ means ${\cal B}(\dblh\rightarrow\ell_\alpha^+\ell_\beta^+) {\cal B}(\dblh\rightarrow\ell_\gamma^+\ell_\delta^+)$ summed over all possible flavor combinations.}
  \end{center}
\end{figure*}

\section*{Acknowledgements}
We congratulate our colleagues in the CERN accelerator departments for the excellent performance of the LHC machine. We thank the technical and administrative staff at CERN and other CMS institutes, and acknowledge support from: FMSR (Austria); FNRS and FWO (Belgium); CNPq, CAPES, FAPERJ, and FAPESP (Brazil); MES (Bulgaria); CERN; CAS, MoST, and NSFC (China); COLCIENCIAS (Colombia); MSES (Croatia); RPF (Cyprus); MoER, SF0690030s09 and ERDF (Estonia); Academy of Finland, MEC, and HIP (Finland); CEA and CNRS/IN2P3 (France); BMBF, DFG, and HGF (Germany); GSRT (Greece); OTKA and NKTH (Hungary); DAE and DST (India); IPM (Iran); SFI (Ireland); INFN (Italy); NRF and WCU (Korea); LAS (Lithuania); CINVESTAV, CONACYT, SEP, and UASLP-FAI (Mexico); MSI (New Zealand); PAEC (Pakistan); MSHE and NSC (Poland); FCT (Portugal); JINR (Armenia, Belarus, Georgia, Ukraine, Uzbekistan); MON, RosAtom, RAS and RFBR (Russia); MSTD (Serbia); MICINN and CPAN (Spain); Swiss Funding Agencies (Switzerland); NSC (Taipei); TUBITAK and TAEK (Turkey); STFC (United Kingdom); DOE and NSF (USA).
Individuals have received support from the Marie-Curie programme and the European Research Council (European Union); the Leventis Foundation; the A. P. Sloan Foundation; the Alexander von Humboldt Foundation; the Belgian Federal Science Policy Office; the Fonds pour la Formation \`a la Recherche dans l'Industrie et dans l'Agriculture (FRIA-Belgium); the Agentschap voor Innovatie door Wetenschap en Technologie (IWT-Belgium); the Council of Science and Industrial Research, India; and the HOMING PLUS programme of Foundation for Polish Science, cofinanced from European Union, Regional Development Fund.

\bibliography{auto_generated}   

\cleardoublepage \appendix\section{The CMS Collaboration \label{app:collab}}\begin{sloppypar}\hyphenpenalty=5000\widowpenalty=500\clubpenalty=5000\input{HIG-12-005-authorlist.tex}\end{sloppypar}
\end{document}

%% file: HIG-12-005-authorlist.tex
\textbf{Yerevan Physics Institute,  Yerevan,  Armenia}\\*[0pt]
S.~Chatrchyan, V.~Khachatryan, A.M.~Sirunyan, A.~Tumasyan
\vskip\cmsinstskip
\textbf{Institut f\"{u}r Hochenergiephysik der OeAW,  Wien,  Austria}\\*[0pt]
W.~Adam, T.~Bergauer, M.~Dragicevic, J.~Er\"{o}, C.~Fabjan\cmsAuthorMark{1}, M.~Friedl, R.~Fr\"{u}hwirth\cmsAuthorMark{1}, V.M.~Ghete, J.~Hammer, N.~H\"{o}rmann, J.~Hrubec, M.~Jeitler\cmsAuthorMark{1}, W.~Kiesenhofer, V.~Kn\"{u}nz, M.~Krammer\cmsAuthorMark{1}, D.~Liko, I.~Mikulec, M.~Pernicka$^{\textrm{\dag}}$, B.~Rahbaran, C.~Rohringer, H.~Rohringer, R.~Sch\"{o}fbeck, J.~Strauss, A.~Taurok, P.~Wagner, W.~Waltenberger, G.~Walzel, E.~Widl, C.-E.~Wulz\cmsAuthorMark{1}
% ...
\vskip\cmsinstskip
\dag:~Deceased\\
1:~~Also at Vienna University of Technology, Vienna, Austria\\
% ...
